\documentclass[aps,apl,twocolumn,showpacs]{revtex4}
\usepackage{graphicx}
\usepackage{hyperref}

\begin{document}

\title{Evaluation of spin diffusion length and spin Hall angle of antiferromagnetic Weyl semimetal Mn$_3$Sn.}

\author{P. K. Muduli$^{1}$}\email{muduli.ps@gmail.com}
\author{T. Higo$^{1}$}
\author{ T. Nishikawa$^{1}$}
\author{D. Qu$^1$}
\author{H. Isshiki$^1$}
\author{K. Kondou$^{2}$}
\author{D. Nishio-Hamane$^{1}$}
\author{S. Nakatsuji$^1$}
\author{YoshiChika Otani$^{1,2}$}\email{yotani@issp.u-tokyo.ac.jp} \affiliation{ $^1$Institute for Solid State Physics, University
of Tokyo, Kashiwa 277-8581, Japan} \affiliation{ $^2$Center for
Emergent Matter Science, RIKEN, 2-1 Hirosawa, Wako 351-0198,
Japan}

\date{\today}
\begin{abstract}

Antiferromagnetic Weyl semimetal Mn$_3$Sn has shown to generate
strong intrinsic anomalous Hall effect (AHE) at room temperature,
due to large momentum-space Berry curvature from the time-reversal
symmetry breaking electronic bands of the Kagome planes. This
prompts us to investigate intrinsic spin Hall effect, a transverse
phenomenon with identical origin as the intrinsic AHE. We report
inverse spin Hall effect experiments in \emph{nanocrystalline}
Mn$_3$Sn nanowires at room temperature using spin absorption
method which enables us to quantitatively derive both the spin
diffusion length and the spin Hall angle in the same device. We
observed clear absorption of the spin current in the Mn$_3$Sn
nanowires when kept in contact with the spin transport channel of
a lateral spin-valve device. We estimate spin diffusion length
$\lambda_{s(Mn_3Sn)}$ $\sim$0.75 $\pm$0.67 nm from the comparison
of spin signal of an identical reference lateral spin valve
without Mn$_3$Sn nanowire. From inverse spin Hall measurements, we
evaluate spin Hall angle $\theta_{SH}$ $\sim$5.3 $\pm$ 2.4 $\%$
and spin Hall conductivity $\sigma_{SH}$ $\sim$46.9 $\pm$ 3.4
($\hbar/e$) ($\Omega$ cm)$^{-1}$. The estimated spin Hall
conductivity agrees with both in sign and magnitude to the
theoretically predicted intrinsic $\sigma_{SH}^{int}$ $\sim$36-96
($\hbar/e$) ($\Omega$ cm)$^{-1}$. We also observed anomalous Hall
effect at room temperature in nano-Hall bars prepared at the same
time as the spin Hall devices. Large anomalous Hall conductivity
along with adequate spin Hall conductivity makes Mn$_3$Sn a
promising material for ultrafast and ultrahigh-density spintronics
devices.

\end{abstract}
\pacs{xx.xx}
\maketitle
\clearpage
\section{Introduction}

Next generation of ultra-fast and ultra-low-power spintronic
devices will be ideally mass-less and dissipation-less. Therefore,
antiferromagnetic materials with topological properties are most
desirable for future spintronics devices. Antiferromagnetic
materials are expected to overtake ferromagnetic materials in
future spintronics devices due to their higher intrinsic
excitation frequency in terahertz (THz) timescale, immunity
against external field perturbations and zero net magnetization
\cite{baltz,smejjal}. Many antiferromagnetic materials have
recently been found to exhibit either topologically protected
massless Dirac or Weyl quasiparticles in their band structure or
topologically non-trivial real-space spin textures. These exotic
antiferromagnets have led to a new area of research called
\emph{topological antiferromagnetic spintronics
}\cite{smejjal-natphy}. So far two antiferromagnets CuMnAs and
Mn$_2$Au exhibiting current-induced N\'{e}el spin-orbit torque are
the prime materials in antiferromagnetic spintronics which have
already led to working devices \cite{Barthem,
wadley,Zelezny-natphy}. Over the past years many other
antiferromagnets like SrMnBi$_2$\cite{Park},
EuMnBi$_2$\cite{Masuda}, BaFe$_2$As$_2$\cite{Richard},
YbMnBi$_2$\cite{Awang}, GdPtBi \cite{muller,Hirschberger},
FeSe\cite{ZFwang}, NdSb\cite{Wakeham},
Eu$_2$Ir$_2$O$_7$\cite{Sushkov}, etc., have emerged which may
enrich topological antiferromagnetic spintronics further.
Noncollinear antiferromagnet Mn$_3X$ ($X$ = Ge, Sn, Ga, Ir, Rh and
Pt) series have been attracting considerable interest lately due
to accidental discovery of the large anomalous Hall effect (AHE)
comparable in magnitude to that of ferromagnets \cite{chen,
nakatsuji, nayak, manna, Kiyohara}. Usually AHE is not realized in
ordinary collinear antiferromagnets, however, recent theoretical
and experimental investigations in chiral antiferromagnets reveal
that a large AHE is possible for non-vanishing Berry phase which
acts as a fictitious magnetic field in momentum space
\cite{chen,kubler-el}.

Here we focus particularly on Mn$_3$Sn which involve both Weyl
physics \cite{kubler, yang} and antiferromagnetism with large
N\'{e}el temperature of $T_N$ $\sim$420 K \cite{Tian}. In Mn$_3$Sn
magneto-geometrical frustration in the Kagome lattice leads to
non-collinear antiferromagnetic order causing Mn moments to lie in
the $ab$-plane (Kagome-plane) with moments aligned at 120$^0$ with
each other. This inverse triangular spin structure carries a very
small net ferromagnetic moment of $\sim$0.002 $\mu_B$/Mn atom,
1000 times smaller than ferromagnets\cite{Brown}. The triangular
spins can be rotated inside the $ab$-plane even with a very weak
magnetic field due to small Kagome-plane magnetic
anisotropy\cite{Nagamiya}. Large anomalous Hall conductivity up to
$\sigma _{xy}^{AHE} $ $\approx$ 120 ($\Omega$ cm)$^{-1}$ has been
observed in Mn$_3$Sn which matches closely to the theoretically
calculated $\sigma _{xy}^{AHE} $ from the integration of Berry
curvature over the Brillouin zone\cite{Guo-prb, nakatsuji}.
Ab-initio band structure calculations \cite{kubler, yang} and
angle-resolved photoemission spectroscopy (ARPES) measurements
\cite{kuroda} have revealed multiple type-2 Weyl points in the
bulk band structure of Mn$_3$Sn. Fermi level has been found to be
as close as $\sim$5 meV to the nearest Weyl node with slightly
extra Mn doping in Mn$_3$Sn\cite{kuroda}. Signatures of chiral
anomaly such as negative longitudinal magnetoresistance and planar
Hall effect has also been observed in Mn$_3$Sn\cite{kuroda}. Large
thermal Hall\cite{Xiaokang}, anomalous Nernst effect \cite{narita,
Ikhlas, Xiaokang},topological Hall effect\cite{Xiaokang-scipost}
and exotic magneto-optical Kerr effect\cite{Higo-nat-photon} has
also been detected in Mn$_3$Sn. Although initial studies on
Mn$_3$Sn was primarily focused on bulk single crystals, recently,
high quality thin films of Mn$_3$Sn showing the exchange-bias
effect \cite{Markou} and large anomalous Hall effect
\cite{THigo,you} have been successfully fabricated and open up
possibility for spintronics device applications.

Spin Hall effect and anomalous Hall effect are analogues phenomena
both originating from the electronic and magnetic structure of the
material. The intrinsic SHE is explained by the spin Berry
curvature which is obtained from Kubo formula, similar to the AHE
\cite{Guo, Nagaosa-rev,Jungwirth,omori}. Therefore, chiral
antiferromagnets are most promising materials for detecting large
spin Hall effect \cite{Sun,Guo-prb,Zhang-prb,zhang}. In direct
spin Hall effect (DSHE) a charge current gives rise to a
transverse spin current which generates spin accumulations with
opposite spin polarization at the reverse sides of a material.
Furthermore, a spin current can also induce a transverse charge
current (voltage drop), in the reciprocal process called the
inverse spin Hall effect (ISHE). The spin-to-charge current
interconversion can be described by, $ \vec J_S  = \frac{e}{\hbar
}\theta _{SH} (\vec J_C \times \vec s)$, where $ \vec J_{S(C)}$ is
the spin(charge) current, $\hbar$ is the reduced Planck's
constant, $e$ is the electronic charge and $\vec s$ denotes the
direction of spin polarization. The conversion efficiency is
characterized by the spin Hall angle, $\theta _{SH}$. Estimation
of spin transport parameters like spin diffusion length
($\lambda_s$), spin Hall angle ($\theta _{SH}$) and spin Hall
conductivity ($\sigma_{SH}$) is indispensable for possible
application of Mn$_3$Sn in spin-orbitronics. Very recently, a
strong SHE was experimentally discovered in another chiral
antiferromagnetic compound IrMn$_3$ and the spin Hall angle up to
$\sim$35$\%$ was observed \cite{Zhang-sciadv}. Theoretical
calculations suggest spin Hall effect in Mn$_3$Sn is strongly
anisotropic and is maximized when charge current $\vec J_C$  and
spin current $\vec J_S$ are inside the
Kagome-plane\cite{Zhang-prb}. Recently, \u{Z}elezn\'{y}\emph{et
al.}\cite{Jakub}  have predicted that spin current in noncollinear
antiferromagnets possess spin components both longitudinal and
transverse to the antiferromagnetic order parameter.
Interestingly, these spin currents are odd under time reversal in
contrast to spin Hall effect spin currents which are even. It is
also expected that the transversal contribution of spin currents
in noncollinear antiferromagnets can be greater than the spin Hall
effect spin currents. These unconventional spin Hall effects in
noncollinear antiferromagnets may open up new avenues in the
understanding of spin Hall effect in antiferromagnets
\cite{huachen,zhang}.

In this paper we use spin absorption in lateral spin valves to
study inverse spin Hall effect in nanocrystalline Mn$_3$Sn
nanowires. The spin absorption method allows us to extract the
spin diffusion length ($\lambda_s$)  and spin Hall angle ($\theta
_{SH}$) in the same device by changing measurement configuration.
In these measurements the antiferromagnetic material is not in
direct contact with the ferromagnetic spin current injector which
avoids exchange-bias effect and make spin absorption method more
reliable way to study spin Hall effect in antiferromagnetic
material. We prepare a set of spin Hall device (SHD), reference
lateral spin-valve and nano-Hall bar on the same substrate to test
both Berry phase induced intrinsic anomalous Hall effect and spin
Hall effect. We estimated $\lambda_s$, $\theta _{SH}$ and
$\sigma_{SH}$ of the nanocrystalline Mn$_3$Sn nanowire at room
temperature and found $\sigma_{SH}$ comparable to theoretical
predictions\cite{Guo-prb, Zhang-prb,zhang}. Large anomalous Hall
conductivity along with moderate spin Hall conductivity adds
further functionality to Mn$_3$Sn for use in topological
antiferromagnetic spintronics.

\section{Experimental details}

Lateral spin-valve devices and nano-Hall bars were fabricated on
Si/SiO$_2$(300 nm) substrate using e-beam lithography in three
steps. In the first step of e-beam lithography a pair of 100 nm
wide and 30 nm thick Py nanowires with distance of 1 $\mu$m were
prepared by e-beam evaporation though a PMMA mask. The Py
deposition was done in an UHV chamber with base pressure lower
than $5\times 10^{-9}$ torr while substrate was kept at 10 $^0$C.
In the second step Mn$_3$Sn nanowires were prepared using direct
current (DC) magnetron sputtering. To avoid side walls in the
nanowire MMA/PMMA bilayer was patterned by e-beam lithography to
form a mask with undercut. Then Mn$_3$Sn was deposited at room
temperature using DC sputtering at rate $\sim$0.2 nm/s with 60 W
of power and 0.3 Pa Ar gas pressure.   We utilized 2 nm Ru seeding
layer as a template to have smooth Mn$_3$Sn surface. After
lift-off the nanowires were annealed in vacuum to 500 $^0$C for 1
hours to achieve stoichiometric Mn$_3$Sn. We successfully
fabricated 11.5 $\mu$m long, 50-70 nm thick and 150 nm wide
nanowires which showed partly metallic electrical transport
properties. Thinner nanowires showed semiconducting temperature
dependence [See supplementary material for details \cite{supp}].
Recently we have used similar post annealing process to achieve
high quality Mn$_3$Sn thin films \cite{THigo}. The structural
characterization of thin films prepared under similar sputtering
conditions to the nanowire and nano-Hall bar were performed using
x-ray diffractometer \cite{supp}. In the third step 100 nm wide
and 100 nm thick Cu was thermally evaporated at rate
$\sim$2{\AA}/s in a separate UHV chamber with base pressure of
$\sim$1.2 $\times$ $10^{-9}$ torr. The surfaces of Py and Mn$_3$Sn
nanowires were in-situ cleaned by Ar-ion milling for 40s before
the Cu deposition. All the devices were capped with 2 nm AlO$_x$
at the end to avoid oxidization. For comparative electrical and
spin transport measurements one set of reference device, spin Hall
device and nano-Hall bar were prepared together on the same
substrate. Reference device and spin Hall devices are identical
lateral spin valves except the Mn$_3$Sn nanowire was inserted in
the middle of two Py electrodes in the later as shown in
Fig.~\ref{Fig2}(a,b). Multiple devices were fabricated on the same
substrate to check reproducibility. All the electrical transport
measurements were done using lock-in technique (173 Hz) in a He-4
flow cryostat.

\begin{center}
\begin{figure}[h]
\begin{tabular}{ll}
  \centering
  \includegraphics[width= 8 cm]{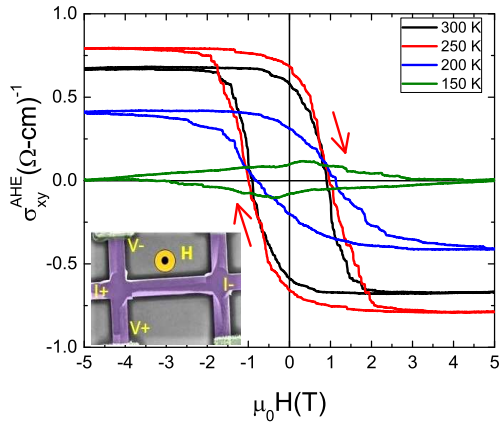}

\end{tabular}
\caption{Magnetic field dependence of the anomalous Hall
conductivity($ \sigma _{xy}^{AHE}  =  - \frac{{\rho _{xy} }}{{\rho
_{xx}^2 }}$) measured at various temperatures between 150 to 300
K. Inset shows SEM picture of a Mn$_3$Sn nano-Hall bar (width =
500 nm, thickness = 70 nm, length =3900 nm) with measurement
configuration.} \label{Fig1}
\end{figure}
\end{center}

\section{Results and discussion}

Growth of Mn$_3$Sn thin films on thermally oxidized Si substrate
strongly depends on the deposition temperature and the choice of
seed layer like Ta, Ru or Pt \cite{Markou,Filippou}. For our
experiments we choose Ru underlayer due to its smaller spin Hall
angle ($\theta_{SH}$ $\approx$0.0056) \cite{wen}. Nanocrystalline
nanowires of Mn$_3$Sn with Ru seed layer were deposited on
thermally oxidized silicon substrate at room temperature and post
annealed ex-situ at 500 $^0$C for crystallization. Bulk Mn$_3$Sn
is known to possess different magnetic structures depending on
small alteration in the chemical composition and growth conditions
\cite{Zimmer}. The intrinsic AHE in Mn$_3$Sn depends sensitively
on the magnetic structure and is maximized in the inverse
triangular spin arrangement\cite{Sung, Xiaokang}. In order to
confirm intrinsic origin of anomalous Hall effect we measure
anomalous Hall conductivity ($ \sigma _{xy}^{AHE} $) in a
nano-Hall bar prepared at the same time as the spin Hall devices.
Fig.~\ref{Fig1} shows magnetic field dependence of anomalous Hall
conductivity, $ \sigma _{xy}^{AHE}  = - \frac{{\rho _{xy}^{AHE}
}}{{\rho _{xx}^2 }}$, at different temperatures after removal of
the high-field linear background (ordinary Hall effect) from the
measured data. Here anomalous Hall resistivity is defined as, $
\rho _{xy}^{AHE}  = \frac{{V_{xy} }}{I}t$, where $V_{xy}$ is the
Hall voltage, $I$ is the applied current and $t$ is the thickness
of Mn$_3$Sn nano-Hall bar. The anomalous Hall conductivity show
clear hysteresis loops with a considerable jump from $-0.6$
($\Omega$ cm)$^{-1}$ (for $\sigma _{xy}^{AHE} (H: + 5T \to 0T)$)
to $+0.6$ ($\Omega$ cm)$^{-1}$ (for $\sigma _{xy}^{AHE} (H: - 5T
\to 0T)$) at room temperature. The anomalous Hall conductivity in
Mn$_3$Sn changes its sign corresponding to the rotation of the Mn
moments of the inverse triangular spin structure. The sign change
in our nano-Hall bar occurs at field $\sim$1 T which is comparable
to that of Mn$_3$Sn polycrystalline thin film \cite{THigo}. The
higher switching field is related to polycrystalline nature of the
Mn$_3$Sn films. The anomalous Hall conductivity was found to
increase up to 250 K and disappear below 150 K as temperature was
lowered. Bulk Mn$_3$Sn is known to undergo phase transition to an
incommensurate spin spiral structure below 275 K which causes
intrinsic contribution of anomalous Hall resistivity disappear
below this temperature \cite{Sung}. This transition temperature is
quite sensitive to synthesis conditions which determine precise
chemical composition (Mn:Sn ratio). The disappearance of AHE below
150 K in our nano-Hall bar is consistent with these observations
in Mn$_3$Sn samples\cite{Sung} indicating the inverse triangular
spin structure at room temperature. In our experiments we
primarily focus on room temperature measurements where inverse
triangular structure-induced Berry curvature seems to be the
dominant contributor to AHE.

\begin{widetext}

\begin{center}
\begin{figure}[h]
\begin{tabular}{ll}
  \centering
  \includegraphics[width= 8 cm]{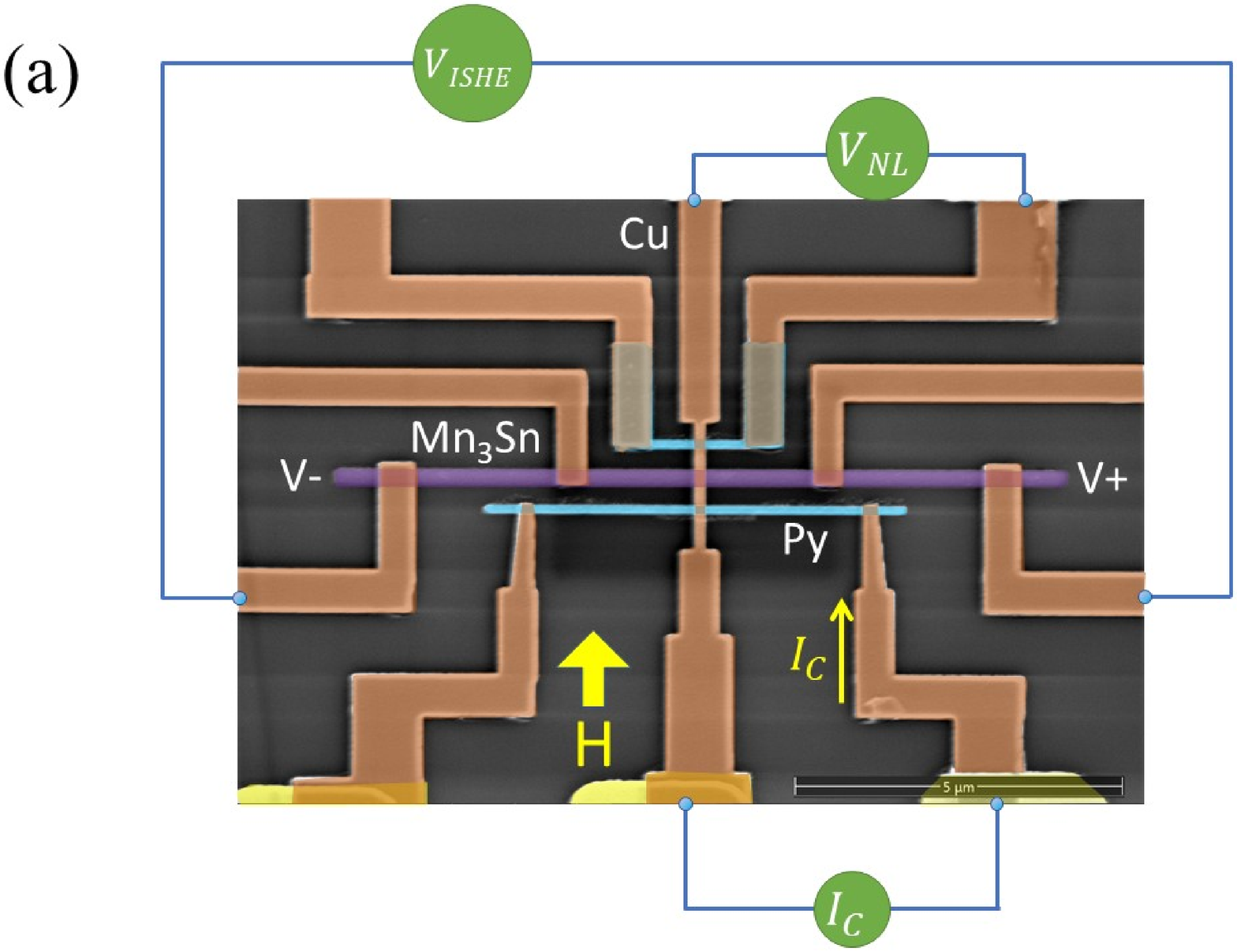}
&

  \includegraphics[width= 7 cm]{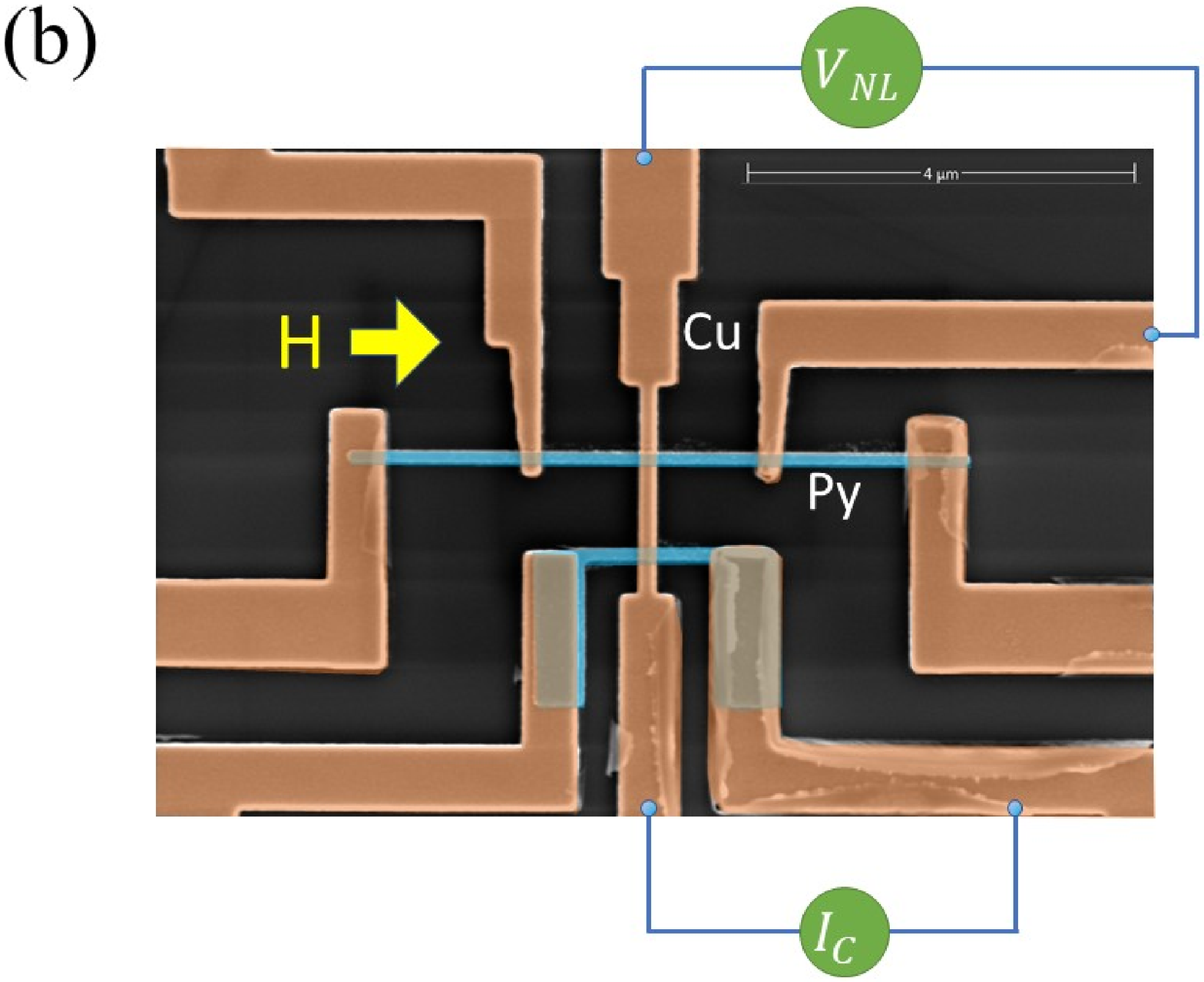}\\

 \centering
 \includegraphics[width= 7 cm]{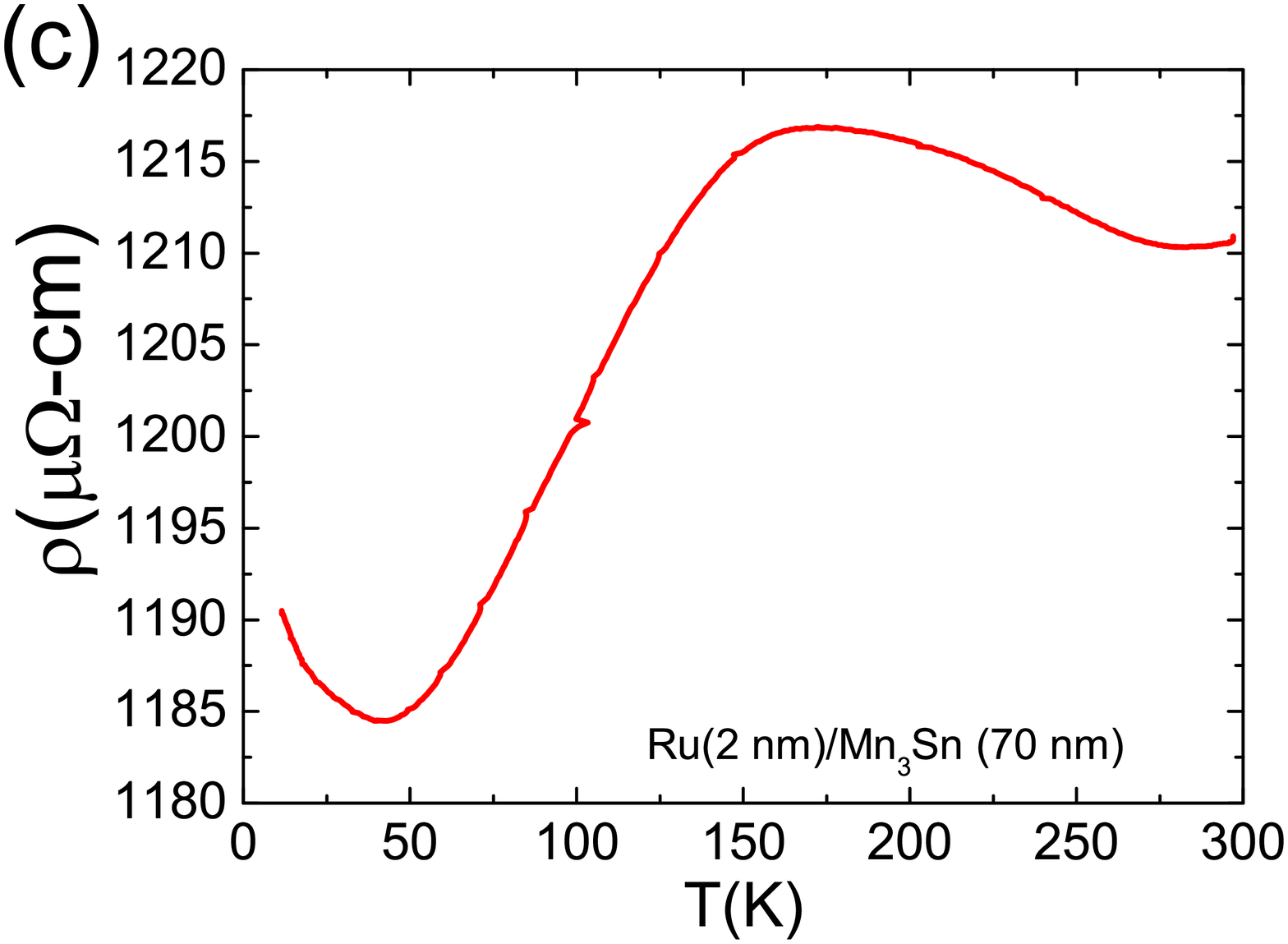}
&

 \includegraphics[width= 7 cm]{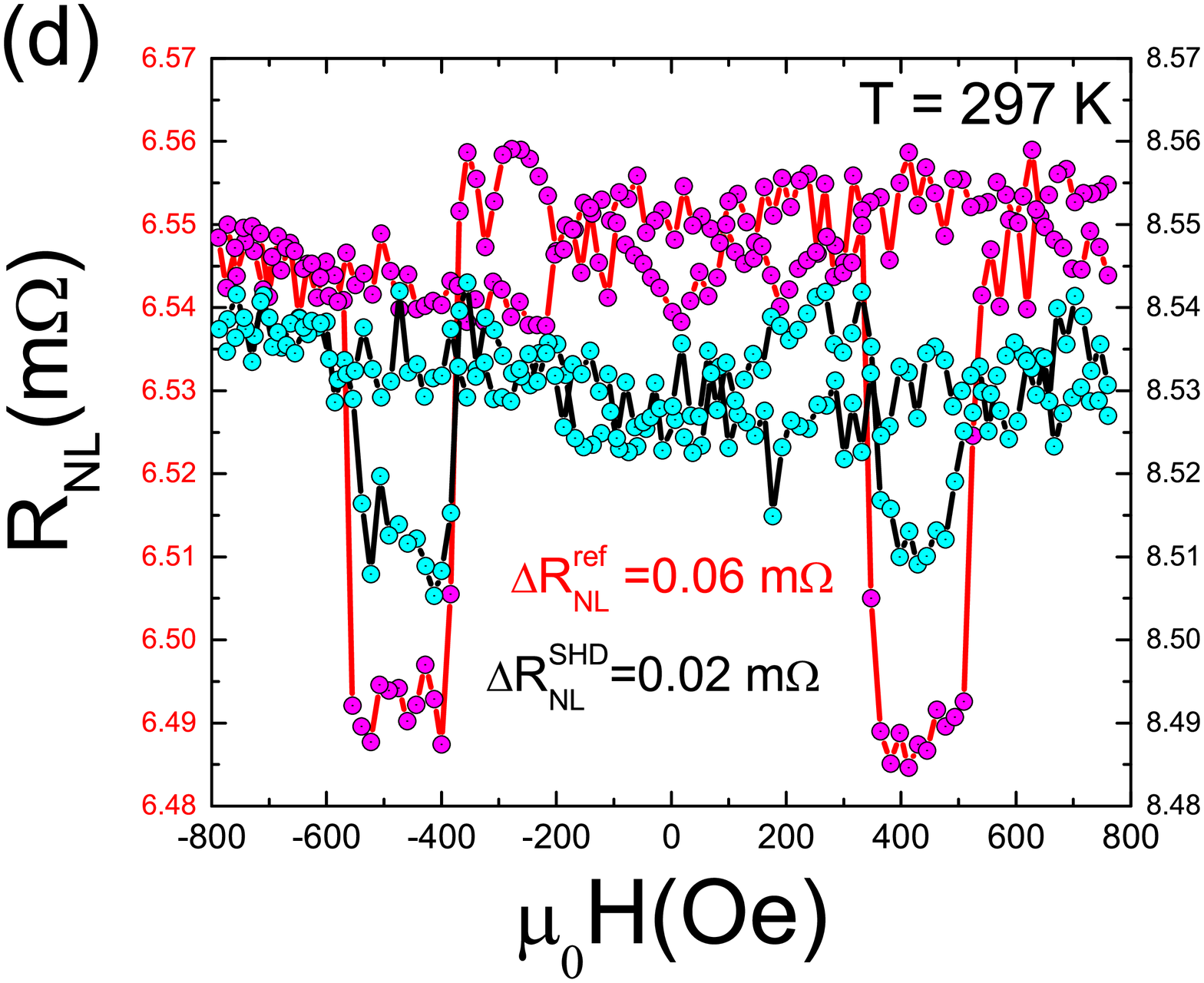}\\

\end{tabular}
\caption{Scanning electron microscopy (SEM) image of (a) spin Hall
device (SHD) and (b) reference lateral spin valve device. The
nonlocal and spin Hall measurement configuration are shown. In
nonlocal measurement configuration  magnetic field ($H$) is
applied inplane along the long easy axis of Py while in spin Hall
measurement configuration $H$ is directed perpendicular (inplane)
to it. (c) Resistivity of Mn$_3$Sn nanowire in the spin Hall
device as a function of temperature.(d) Nonlocal resistance
$R_{NL}$ as a function of magnetic field for reference and spin
Hall device measured at room temperature.} \label{Fig2}
\end{figure}
\end{center}
\end{widetext}
Figure  ~\ref{Fig2}(a) and (b) presents schematic of a spin Hall
and reference lateral spin-valve device, respectively. First, we
affirm the quality of Mn$_3$Sn nanowire from temperature
dependence of resistance measurement (see Supplementary material
for further details). Ideal nanowire is expected to show metallic
temperature dependence with   resistivity $\rho \approx320
\mu\Omega$ cm at room temperature as in bulk Mn$_3$Sn
\cite{nakatsuji}. Fig.~\ref{Fig2}(c) shows temperature dependence
of the resistivity of a Mn$_3$Sn nanowire in the spin Hall device.
The electrical resistivity exhibits a partly metallic behavior
with resistivity $\rho \approx1133 \mu\Omega$ cm at room
temperature. Higher resistivity in the nanowire might be related
to additional electron scattering from the surface as a
consequence of reduced dimension \cite{Fuchs}. Below 50 K an
upturn in the resistivity was observed which is reminiscent of a
spin-glass state.  It is well known that in bulk Mn$_3$Sn a
cluster glass phase appears below 50 K due to spin canting towards
c-axis \cite{nakatsuji,Sung}. Although in bulk single crystals no
resistivity up-turn is observed below 50 K, in nanocrystalline
nanowires blocked spins at the surface can cause Kondo-like
up-turn in the resistivity.

Next, we performed spin absorption experiments in the nonlocal
measurement configuration as shown in Fig.~\ref{Fig2}(a,b).  In
nonlocal spin signal measurements, a nonequilibrium spin
accumulation is created inside Cu spin transport channel by
injecting spin polarized current from one Py electrode into the Cu
channel. The accumulated spin inside Cu diffuses towards the
second Py detector electrode creating pure spin current. The
nonlocal resistance is defined as, $R_{NL} = V_{NL}/I_C$, where
$V_{NL}$ is the nonlocal voltage at the detector and $I_C$ is the
charge current through the injector. When the Mn$_3$Sn nanowire is
kept in contact with the Cu spin transport channel a part of the
spin current is absorbed by it due to lower spin resistance of
Mn$_3$Sn compared to Cu. The spin resistance is a quantity
equivalent of electrical resistance but for spin current and is
defined as, $R_S = \frac{{\rho \lambda_s }}{{(1 - p^2)A}}$, where
$\rho$ is the resistivity, $\lambda_s$ is the spin diffusion
length, $p$ is the spin polarization and $A$ is the area of
cross-section of the spin transport channel. In order to estimate
amount of spin current absorbed by the Mn$_3$Sn nanowire, nonlocal
resistance was measured both in the spin Hall device (Fig.
~\ref{Fig2}(a)) and another reference lateral spin valve device
(Fig. ~\ref{Fig2}(b)). The spin signal $\Delta R_{NL}$ is
expressed as, $\Delta R_{NL}  = R_{NL}^{ \uparrow  \uparrow }  -
R_{NL}^{ \uparrow \downarrow }$, where $R_{NL}^{ \uparrow \uparrow
} (R_{NL}^{ \uparrow \downarrow })$ is the nonlocal resistance
when both injector and detector Py are aligned parallel
(antiparallel) to each other. Fig. ~\ref{Fig2}(d) shows $R_{NL}$
as a function of magnetic field for both spin Hall and reference
lateral spin valve device at room temperature. Smaller $\Delta
R_{NL}$ was observed in the spin Hall device compared to the
reference lateral spin valve suggesting spin current absorption by
Mn$_3$Sn nanowire. Similar spin current absorption was also
observed at all temperature (see Supplementary material for
measurement at 10 K)\cite{supp}. The spin resistance of Mn$_3$Sn
nanowire can be obtained from ratio of spin signals between spin
Hall ($\Delta R_{NL}^{SHD}$) and reference lateral spin valve
($\Delta R_{NL}^{ref}$) device. Assuming one dimensional spin
diffusion and transparent interfaces the ratio of spin signals can
be expressed as \cite{edurne, Sagasta}

\begin{widetext}

\begin{equation}\label{eq1}
\begin{array}{l}
 \frac{{\Delta R_{NL}^{SHD} }}{{\Delta R_{NL}^{ref} }} =  \\
 {\rm{       }}\frac{{\left[ {2Q_{Mn3Sn} [\sinh \left( {\frac{L}{{\lambda _{s(Cu)} }}} \right) + 2Q_{Py} e^{L/\lambda _{s(Cu)} }  + 2Q_{Py}^2 e^{L/\lambda _{s(Cu)} } ]} \right]}}{{\left[ \begin{array}{l}
 \cosh \left( {\frac{L}{{\lambda _{s(Cu)} }}} \right) - \cosh \left( {\frac{{L - 2d}}{{\lambda _{s(Cu)} }}} \right) + 2Q_{Py} \sinh \left( {\frac{d}{{\lambda _{s(Cu)} }}} \right)e^{(L - d)/\lambda _{s(Cu)} }  + 2Q_{Mn3Sn} \sinh \left( {\frac{L}{{\lambda _{s(Cu)} }}} \right) +  \\
 {\rm{                             }}4Q_{Py} Q_{Mn3Sn} e^{L/\lambda _{s(Cu)} }  + 2Q_{Py} \sinh \left( {\frac{{L - d}}{{\lambda _{s(Cu)} }}} \right)e^{d/\lambda _{s(Cu)} }  + 2Q_{Py}^2 e^{L/\lambda _{s(Cu)} }  + 4Q_{Py}^2 Q_{Mn3Sn} e^{L/\lambda _{s(Cu)} }  \\
 \end{array} \right]}} \\.
 \end{array}
\end{equation}

\end{widetext}

Where $Q_{Py(Mn_3Sn)}  = \frac{{R_{Py(Mn_3Sn)} }}{{R_{Cu} }}$,
with $R_{Cu}  = \frac{{\lambda _{s(Cu)} \rho _{Cu} }}{{w_{Cu}
t_{Cu} }}$, $R_{Py} = \frac{{\lambda _{s(Py)} \rho _{Py}
}}{{w_{Py} w_{Cu} (1 - p_{Py}^2 )}}$ and $R_{Mn_3Sn} =
\frac{{\lambda _{s(Mn_3Sn)} \rho _{Mn_3Sn} }}{{w_{Mn_3Sn} w_{Cu}
}}tanh \left( {\frac{{t_{Mn_3Sn} }}{{\lambda _{s(Mn_3Sn)} }}}
\right)$ are the spin resistances of Cu , Py and Mn$_3$Sn
nanowires, respectively. Here $\rho _i$, $\lambda _s(i)$, $p_i$,
$w_i$ and $t_i$ are resistivity, spin diffusion length, spin
polarization, width and thickness of corresponding nanowires,
respectively ($i$ = Py, Cu and Mn$_3$Sn). Here $L$ is the
center-to-center distance between two Py electrodes and $d$ is the
distance of the Mn$_3$Sn nanowire from injector Py electrode which
was determined from SEM image of the device. The spin resistance
values $R_{Cu}$ and $R_{Py}$ were taken from our previous work
\cite{muduli}. With measured value of the ratio $\frac{{\Delta
R_{NL}^{SHD} }}{{\Delta R_{NL}^{ref} }}$, the spin resistance of
Mn$_3$Sn nanowire can be obtained by solving Eq.~\ref{eq1}. Using
$\frac{{\Delta R_{NL}^{SHD} }}{{\Delta R_{NL}^{ref} }}$ =0.33, we
found $R_{Mn_3Sn}$  = 0.2056 $\Omega$. With resistivity of
Mn$_3$Sn nanowire $\rho_{Mn_3Sn} \approx$1133 $\mu\Omega$ cm we
estimate spin diffusion length of Mn$_3$Sn to be
$\lambda_{s(Mn_3Sn)}$ $\sim$0.75 $\pm$0.67 nm at room temperature.
Recently, spin diffusion length has been measured in a variety of
antiferromagnetic Mn alloys like IrMn, FeMn, PtMn and PdMn, etc.,
\cite{WZhang-prl}. Spin diffusion length has been found to be
quite short $\lambda_s$ $\sim$1 nm in all these antiferromagnets.
Our calculated spin diffusion length for Mn$_3$Sn
$\lambda_{s(Mn_3Sn)}$ $\sim$0.75 $\pm$0.67 nm is consistent with
these previous findings. Spin diffusion length also depends
sensitively on the resistivity of the antiferromagnetic metal and
can be further tuned with $\rho_{Mn3Sn}$ of the nanowire
\cite{baltz}.

\begin{center}
\begin{figure}[h]
\begin{tabular}{ll}
  \centering
  \includegraphics[width= 8 cm]{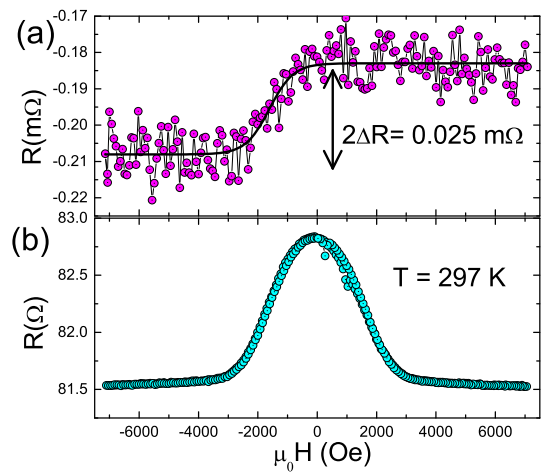}

\end{tabular}
\caption{(a) Magnetic field dependence of Inverse spin Hall
resistance $R_{ISHE}$ measured at room temperature. The inverse
spin Hall resistance $R_{ISHE}$  saturates above $\pm$3000 Oe when
Py magnetization is aligned along applied magnetic field. The
difference between positive and negative saturated $R_{ISHE}$ is
called inverse spin Hall signal $ 2\Delta R_{ISHE}$. (b)
Anisotropic magnetoresistance (AMR) of the Py nanowire in the same
device. The spin polarization direction $\vec{s}$ of the spin
current $\vec{I_s}$ is determined by the magnetization direction
of Py injector electrode. } \label{Fig3}
\end{figure}
\end{center}

After estimating spin resistance of Mn$_3$Sn nanowire we switch
the measurement configuration to the spin Hall measurement as
shown in Fig. ~\ref{Fig2}(a).  In this measurement configuration
magnetic field ($H$) is applied perpendicular and inplane to the
long easy-axis of the Py nanowire as spin current $\vec I_S$,
charge current $\vec I_C$ and spin polarization $\vec s$ are
mutually orthogonal to each other as enforced by the equation,
$\vec I_S = \frac{e}{\hbar }\theta _{SH} (\vec I_C \times \vec
s)$. Fig. ~\ref{Fig3}(b) shows two-terminal resistance ($R-H$) of
the Py nanowire as a function of magnetic field. This anisotropic
magnetoresistance (AMR) or $R-H$ measurement reflects
magnetization direction of the Py with respect to applied magnetic
field $H$. The resistance is minimum when the magnetization of Py
nanowire is aligned along the applied field direction. Due to
shape anisotropy the magnetization of Py nanowire is aligned along
the long easy-axis in zero magnetic field. From Fig.
~\ref{Fig3}(b) Py nanowire can be seen to saturate along the hard
axis when $H >$ $\pm$3000 Oe.

Alike previous measurement, pure spin current is created inside Cu
channel by injecting charge current $I_C$ through one of the Py
electrode which is partially absorbed by the Mn$_3$Sn nanowire.
Due to inverse spin Hall effect a charge current is produced
inside the Mn$_3$Sn nanowire orthogonal to both the spin current
$\vec I_S$ and spin polarization  $\vec s$ direction. In open
circuit condition a voltage drop $V_{ISHE}$ is generated along the
Mn$_3$Sn nanowire due to this charge current. The inverse spin
Hall resistance is defined as, $R_{ISHE} =V_{ISHE}/I_C$, where
$I_C$ is the injected charge current through the Py electrode.
When Py magnetization is switched with applied magnetic field the
orientation of spin polarization changes causing opposite
$R_{ISHE}$ (or $V_{ISHE}$). The difference of the two $R_{ISHE}$
yields twice the inverse spin Hall effect signal $2\Delta
R_{ISHE}$. Fig. ~\ref{Fig3}(a) shows $R_{ISHE}$ as a function of
magnetic field at room temperature. The measurement was done with
injector current $I_C$ = 500 $\mu$A. The inverse spin Hall
resistance can be seen to saturate above $\pm$3000 Oe when Py
electrodes are aligned along the applied magnetic field. We found
a small $2\Delta R_{ISHE}  = 0.025 m\Omega$ from the difference
between $R_{ISHE}$ for the two spin polarization direction. The
spin Hall resistivity can be calculated from the equation
\cite{edurne,Sagasta, Morota}

\begin{equation}\label{eq2}
\rho _{SH}  =  - \frac{{w_{Mn3Sn} }}{x_{sh}}\left( {\frac{{I_C
}}{{\bar I_S }}} \right)\Delta R_{ISHE}.
\end{equation}

Where $x_{sh}$ is the shunting factor which takes into account the
charge current in the Mn$_3$Sn that is shunted through the more
conductive Cu nanowire on top. The shunting factor $x_{sh}$ can be
calculated numerically using finite element method with COMSOL
software (see Supplementary material for details). Here $\bar I_S$
is the effective spin current that contributes to the ISHE voltage
in Mn$_3$Sn and can be expressed as \cite{edurne,Sagasta}
\begin{widetext}
\begin{equation}\label{eq3}
\begin{array}{l}
 \frac{{\bar I_s }}{{I_c }} =  \\
 {\rm{        }}\frac{{\lambda _{s(Mn3Sn)} }}{{t_{Mn3Sn} }}\frac{{(1 - e^{ - t_{Mn3Sn} /\lambda _{s(Mn3Sn)} } )^2 }}{{(1 - e^{ - 2t_{Mn3Sn} /\lambda _{s(Mn3Sn)} } )}} \times  \\
 {\rm{                }}\frac{{\left[ {2\alpha _{Py} [Q_{Py} \sinh \left( {\frac{{L - d}}{{\lambda _{s(Cu)} }}} \right) + Q_{Py}^2 e^{\frac{{L - d}}{{\lambda _{s(Cu)} }}} ]} \right]}}{{\left[ \begin{array}{l}
 \cosh \left( {\frac{L}{{\lambda _{s(Cu)} }}} \right) - \cosh \left( {\frac{{L - 2d}}{{\lambda _{s(Cu)} }}} \right) + 2Q_{Py} \sinh \left( {\frac{d}{{\lambda _{s(Cu)} }}} \right)e^{\frac{{L - d}}{{\lambda _{s(Cu)} }}}  + 2Q_{Mn3Sn} \sinh \left( {\frac{L}{{\lambda _{s(Cu)} }}} \right) + 4Q_{Py} Q_{Mn3Sn} e^{\frac{L}{{\lambda _{s(Cu)} }}}  +  \\
 {\rm{                            }}2Q_{Py} e^{\frac{d}{{\lambda _{s(Cu)} }}} \sinh \left( {\frac{{L - d}}{{\lambda _{s(Cu)} }}} \right) + 2Q_{Py} ^2 e^{\frac{L}{{\lambda _{s(Cu)} }}}  + 4Q_{Py} ^2 Q_{Mn3Sn} e^{\frac{L}{{\lambda _{s(Cu)} }}}  \\
 \end{array} \right]}} \\.
 \end{array}
\end{equation}
\end{widetext}

Using Eqs. ~\ref{eq2} and ~\ref{eq3}, we found $ \rho _{SH}$ = -
60.33 $\pm$ 26.48 $\mu \Omega$  cm. The spin Hall angle given by
the ratio of spin Hall resistivity against electrical resistivity
can be calculated as, $\theta _{SH}  =  - \frac{{\rho _{SH}
}}{{\rho _{Mn3Sn} }}$ , where $\rho _{Mn3Sn}  = 1133 \mu \Omega$
cm is resistivity of the Mn$_3$Sn nanowire\cite{note1}. We
estimate $\theta _{SH}  = 0.053 \pm 0.024$ for our Mn$_3$Sn
nanowire which is comparable to $\theta _{SH}$ of $4d$ and $5d$
transition heavy metals determined by similar spin absorption
method \cite{Morota,niimi}. The spin Hall resistivity is related
to spin hall conductivity as, $\sigma _{SH} \approx  - \frac{{\rho
_{SH} }}{{\rho ^2 _{Mn3Sn} }}$. We found $\sigma _{SH}  = 46.99
\pm 3.42\left( {\frac{\hbar }{e}} \right)(\Omega cm)^{ - 1}$.
Recent theoretical investigations have predicted a positive sign
of intrinsic spin Hall conductivity for Mn$_3$Sn and magnitude in
the range $\sigma_{SH}^{int}$ $\sim$36-96 ($\hbar/e$) ($\Omega$
cm)$^{-1}$ \cite{Zhang-prb, Guo-prb,zhang}. Our estimated
$\sigma_{SH}$ is with in the range of these theoretical
predictions based on Berry curvature calculations. Recently, Mn-Sn
alloy films has shown to exhibit a large spin Hall angle
\cite{Qu}. However, spin Hall angle was found to be negative in
that case. Positive spin Hall angle observed in our case suggest
measured inverse spin Hall signal does not originate from impurity
phases. In Eq.~\ref{eq2}, we assume that all the absorbed spin
current is converted to electrical voltage via inverse spin Hall
effect. However, there might be other sources of spin memory loss
related to magnetization dynamics inside the antiferromagnet. It
is quite challenging to estimate this exactly and is the main
bottleneck of spin absorption method for antiferromagnetic
material.

Nanowire used in this work are nanocrystalline and contain
randomly oriented Kagome planes\cite{supp}. As per theoretical
predictions in order to observe large $\sigma _{SH}$, one should
set the charge and spin currents inside the Kagome plane
\cite{Zhang-prb}. Our results are comparable to structurally
similar material IrMn$_3$ which also shows spin diffusion length
less than 1 nm and spin Hall angle vary between $\theta_{SH}$ =
3-12$\%$ depending of composition $x$ of Ir$_{1-x}$Mn$_x$
\cite{Zhang-sciadv}. In IrMn$_3$ spin Hall effect is believed to
originate from two sources, (i) bulk spin-orbit coupling of
IrMn$_3$ and (ii) the triangular spin structure also gives rise to
an intrinsic spin Hall effect that is large and strongly depends
on the crystallographic orientation of the epitaxial film. Highly
oriented Mn$_3$Sn nanowires may be needed in order to observe
theoretically predicted odd spin currents related to inverse
triangular spin structure\cite{Jakub} . Recently, we have observed
that the spin Hall angle in single crystal Mn$_3$Sn can be
switched with the direction of the staggered moment in the inverse
triangular spin structure \cite{Kimata}.

\section{Conclusions}

To summarize we have investigated inverse spin Hall effect in
nanocrystalline Mn$_3$Sn nanowires by spin absorption method. We
have estimated positive spin Hall angle $\theta_{SH}$ $\sim$5.3
$\pm$ 2.4 $\%$ ,  spin diffusion length $\lambda_{s(Mn_3Sn)}$
$\sim$0.75 $\pm$0.67 nm  and spin Hall conductivity $\sigma_{SH}$
$\sim$46.9 $\pm$ 3.4 ($\hbar/e$) ($\Omega$ cm)$^{-1}$. This
$\sigma_{SH}$ is quite close to theoretically predicted intrinsic
spin Hall conductivity of Mn$_3$Sn suggesting intrinsic origin of
spin Hall effect in our nanowires\cite{Guo-prb,Zhang-prb,zhang}.
The spin Hall and anomalous Hall conductivity can be further
improved by resistivity, strain\cite{Liu} and chemical
tuning\cite{Ikhlas} of the present Mn$_3$Sn nanowire. These
results are obtained in a nanocrystalline nanowire which may have
random orientations of Kagome planes. Further structural
characterization is required on the nanowires to understand
detailed spin structure and orientation of Kagome planes. With
highly oriented nanowires it might be possible to investigate
theoretically predicted novel odd spin currents resulting from the
noncollinear magnetic structure. Reasonably large spin Hall
conductivity of Mn$_3$Sn comparable to other Mn-alloy based
antiferromagnets \cite{WZhang-prl}suggest it can be used as an
efficient spin current detector in nanometer-sized
antiferromagnetic devices. Inclusion of spin Hall effect further
broadens great potential of Mn$_3$Sn in the antiferromagnetic
spintronic devices that could enable spin-based operations at the
ultimate THz frequencies.

\noindent \textbf{Acknowledgments}\\
We thank  Dr Y. Niimi, Dr E. Sagasta, Prof F\`{e}lix Casanova and
Prof S. P. Dash for helpful discussions. This work was supported
by CREST(JPMJCR15Q5), Grant-in-Aid for Scientific Research on
Innovative Area (Grant No. 26103002, 15H05882 and 15H05883) from
the Ministry of Education, Culture, Sports, Science, and
Technology of Japan. Lithography facilities provided by Dr. T.
Nakamura and Prof. S. Katsumoto is gratefully acknowledged.

\newpage

\begin{widetext}
\widetext
\begin{center}
\textbf{\large \underline{Supplementary Material}\\
Evaluation of spin diffusion length and spin Hall angle of
antiferromagnetic Weyl semimetal Mn$_3$Sn.}

P. K. Muduli$^{1}$,T. Higo$^{1}$, T. Nishikawa$^{1}$, Danru
Qu$^1$, H. Isshiki$^1$, K. Kondou$^{2}$, D. Nishio-Hamane$^{1}$,
S. Nakatsuji$^1$, YoshiChika Otani$^{1,2}$
\\
\emph{ $^1$Institute for Solid State Physics, University of Tokyo,
Kashiwa 277-8581, Japan}
\\
\emph{ $^2$Center for Emergent Matter Science, RIKEN, 2-1
Hirosawa, Wako 351-0198, Japan}

\end{center}

\setcounter{page}{1}
\setcounter{section}{0}

\renewcommand{\thefigure}{S\arabic{figure}}
\setcounter{figure}{0}

\renewcommand\theequation{S\arabic{equation}}
\setcounter{equation}{0}

\section{Quality of M\lowercase{n$_3$}S\lowercase{n} nanowire}

Fabrication of high-quality nanowires with thickness $\sim$10-50
nm and width $\sim$100-200 nm are experimentally quite
challenging. In this manuscript we follow a bottom up approach for
device fabrication. The Mn$_3$Sn nanowires were deposited by DC
sputtering method on MMA/PMMA mask prepared by e-beam lithography.
Ideally thinner nanowires with thickness comparable to its' spin
diffusion length are most suitable for spin-absorption
experiments. In our experiments we used thicker $\sim$70 nm
nanowires as thinner ($<$50 nm) nanowires were found to show
semiconducting-like $R(T)$ probably due to oxidation in
atmosphere.

\begin{center}
\begin{figure}[h]
\begin{tabular}{ll}
  \centering
  \includegraphics[width= 14 cm]{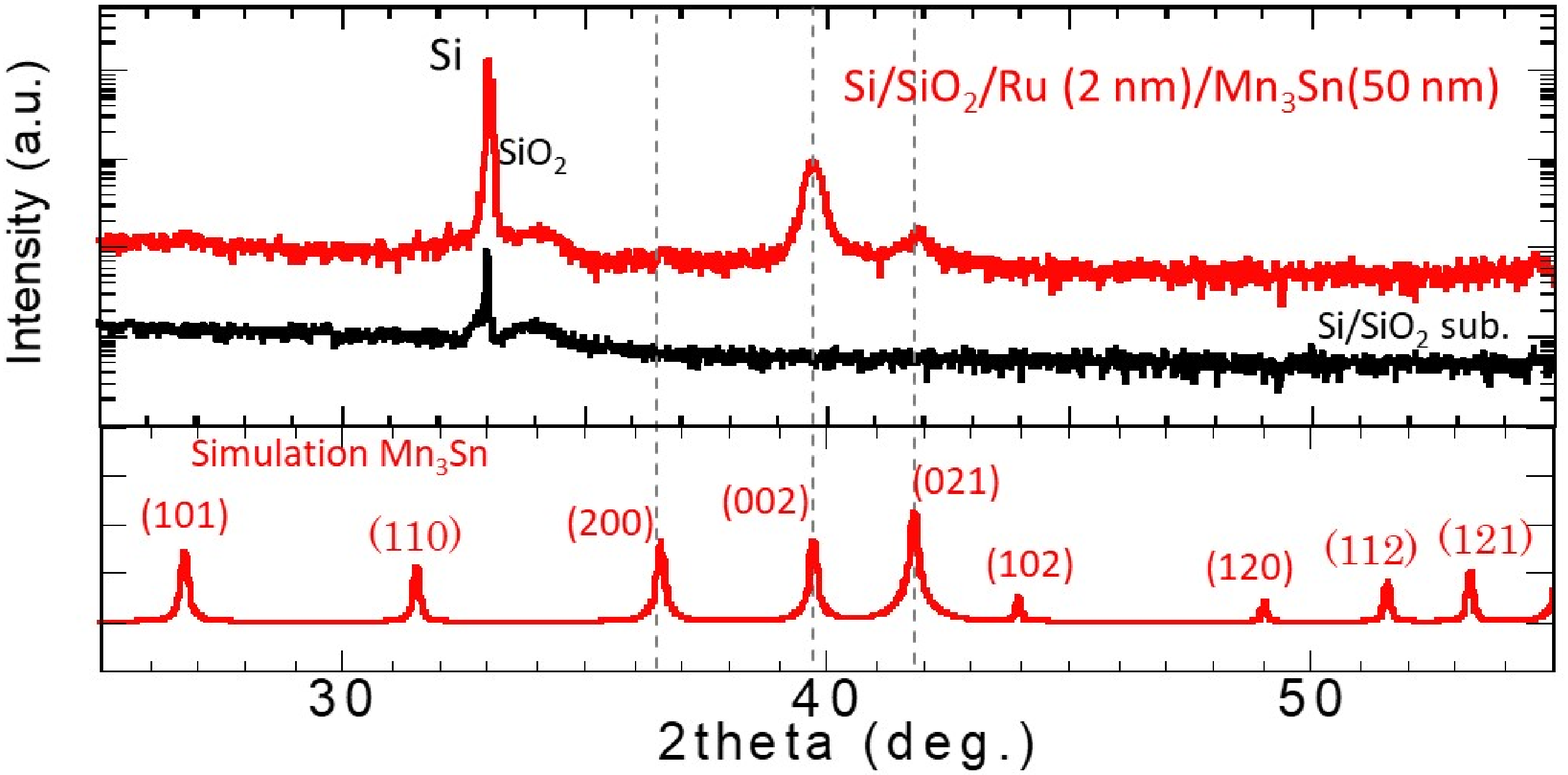}

\end{tabular}
\caption{X-ray diffraction scan of 50 nm Mn$_3$Sn thin film with 2
nm Ru seed layer (red) and bare Si/SiO$_2$ substrate (black).
Bottom panel shows simulated x-ray diffraction pattern of Mn$_3$Sn
with all possible peaks indexed.} \label{FigS1}
\end{figure}
\end{center}

Besides nanowires, Mn$_3$Sn thin films were also fabricated under
identical sputtering condition for structural characterization.
Note that thin films were annealed insitu while nanowires were
annealed ex-situ after the lift-off process. Therefore, nanowires
might be more disordered compared to thin films. Fig. S1 shows
x-ray diffraction spectra of a 50 nm thick Mn$_3$Sn film with 2 nm
Ru seed layer. For comparison x-ray diffraction spectra of one
bare Si/SiO$_2$ substrate without Mn$_3$Sn film is also plotted.
Bottom panel shows simulated x-ray diffraction spectra of the
hexagonal Mn$_3$Sn structure with all crystallographic directions
indexed. Broad peaks corresponding to (002), (021) and (200) plane
of Mn$_3$Sn were detected confirming polycrystalline nature of the
films. A more intense (002) peak was observed suggesting
preferential hexagonal (0001)-axis oriented texture of the films.

\begin{center}
\begin{figure}[h]
\begin{tabular}{ll}
  \centering
  \includegraphics[width= 7 cm]{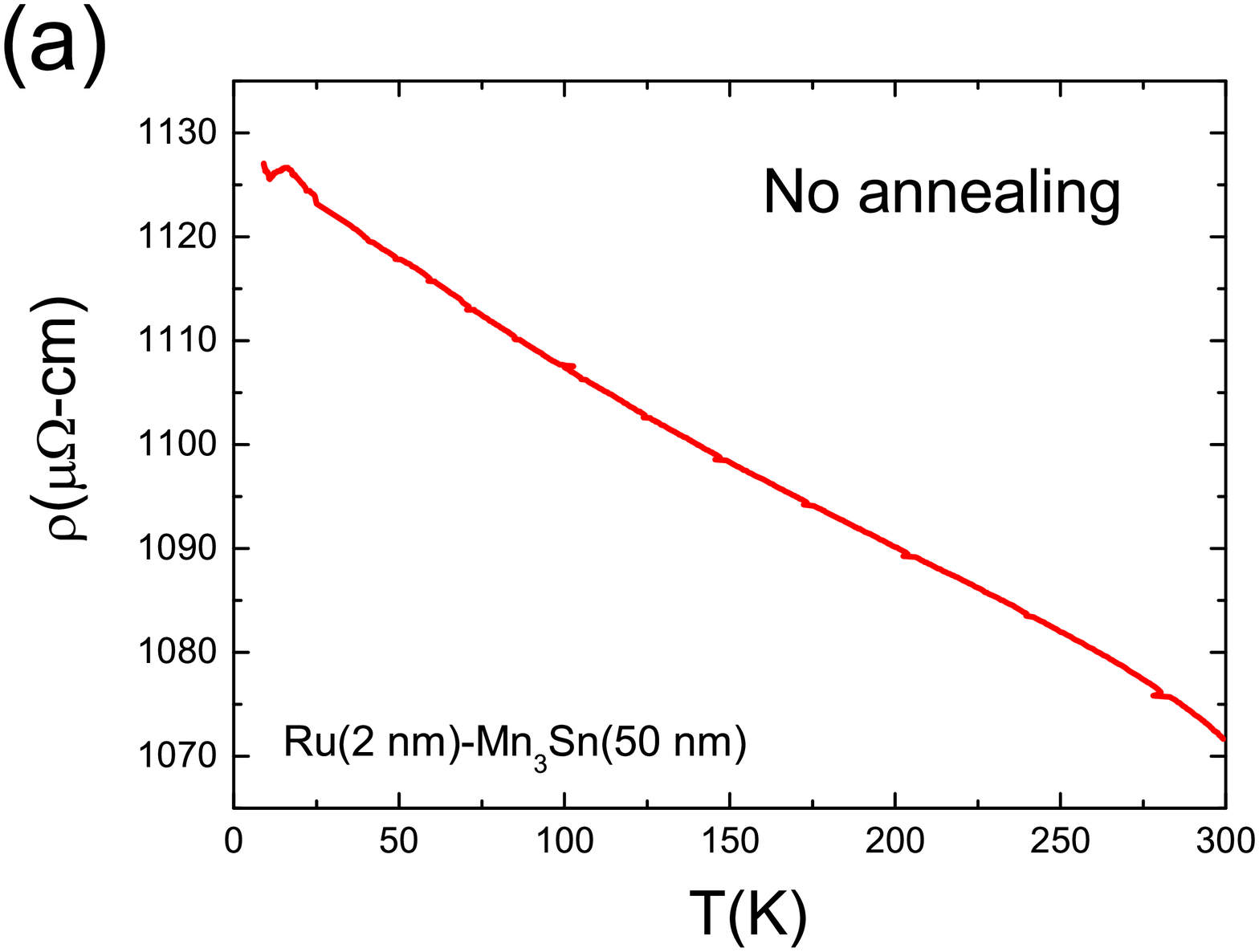}
&

  \includegraphics[width= 7 cm]{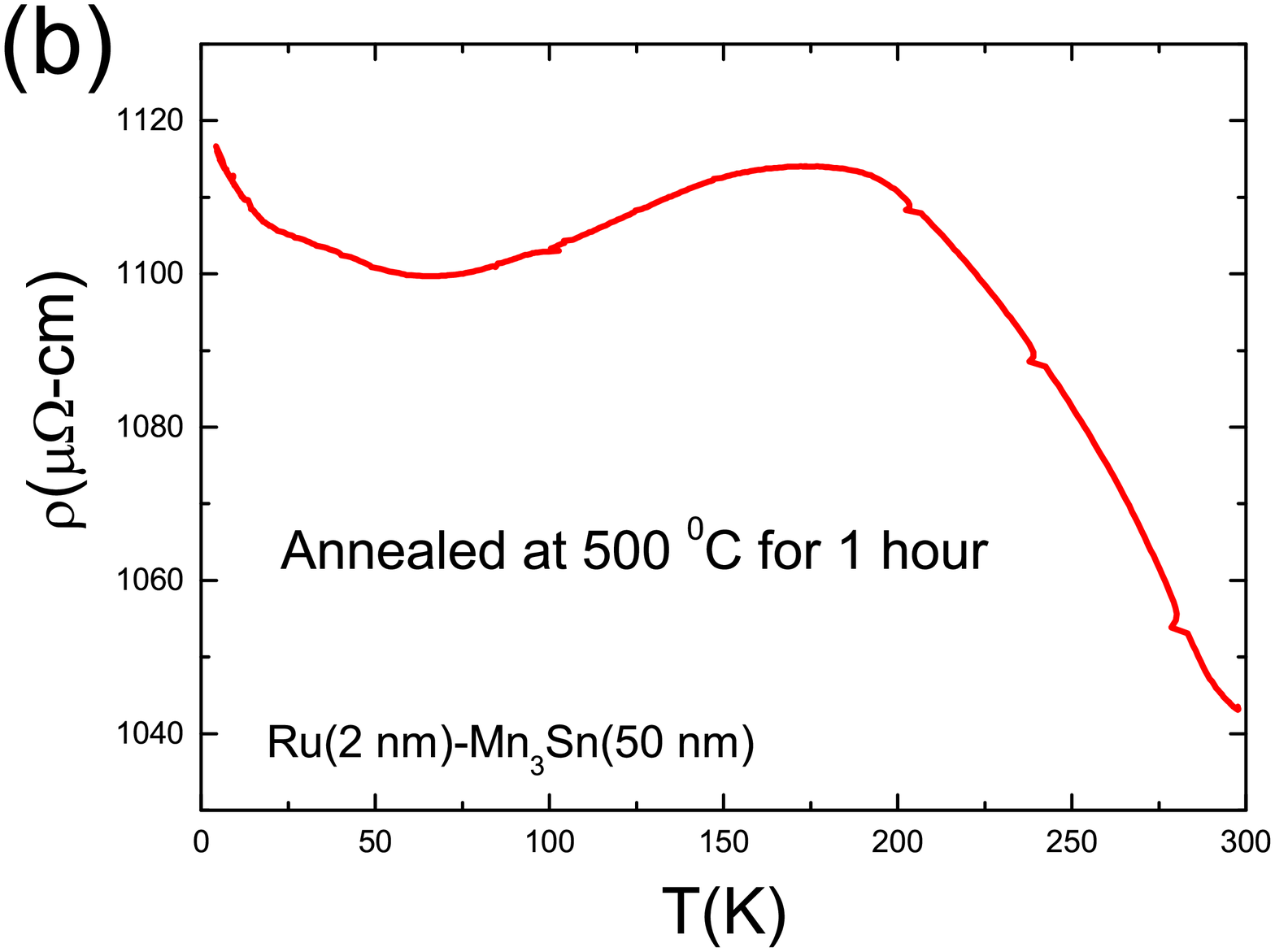}\\

 \centering
 \includegraphics[width= 7 cm]{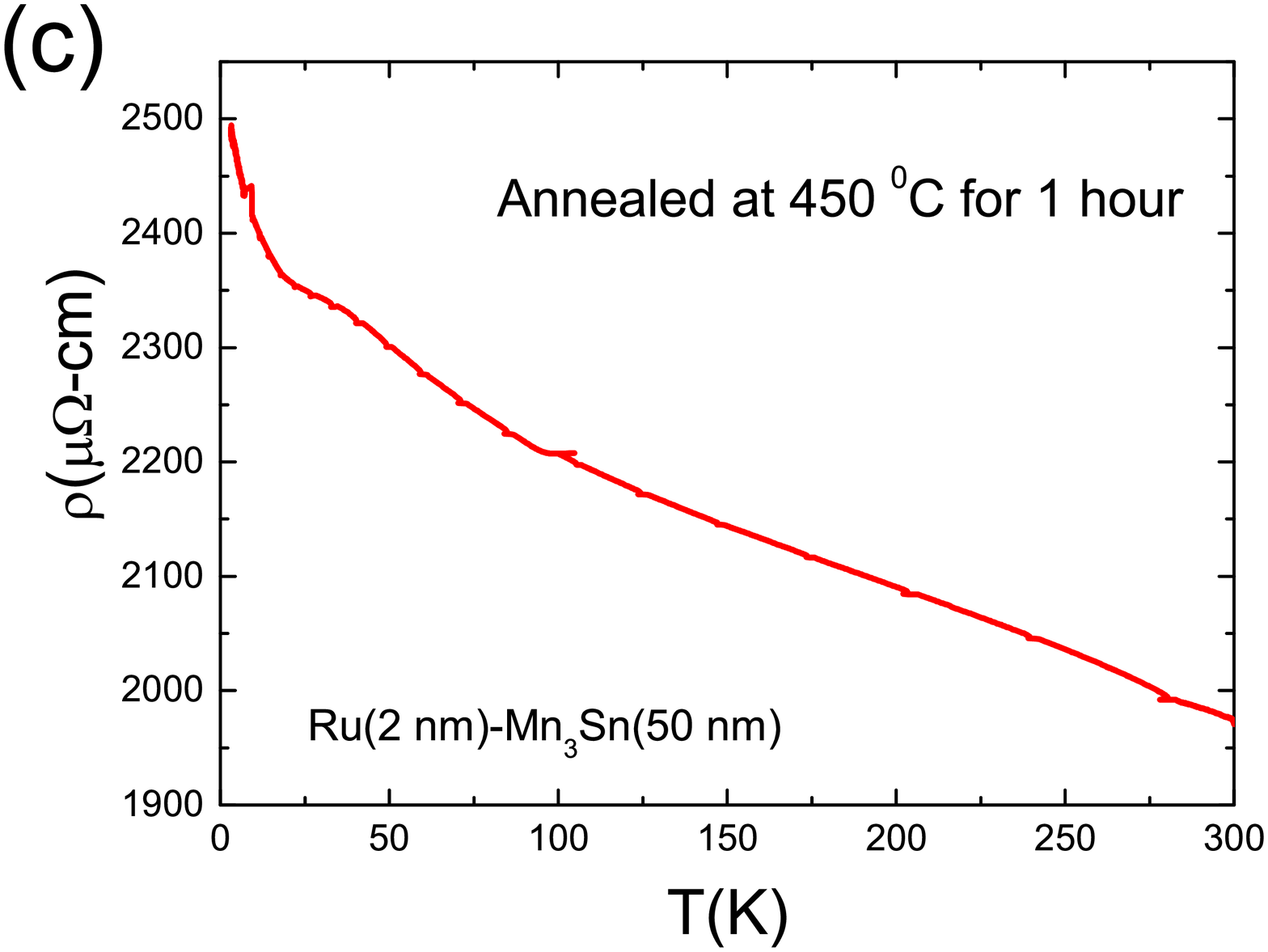}
&

 \includegraphics[width= 7 cm]{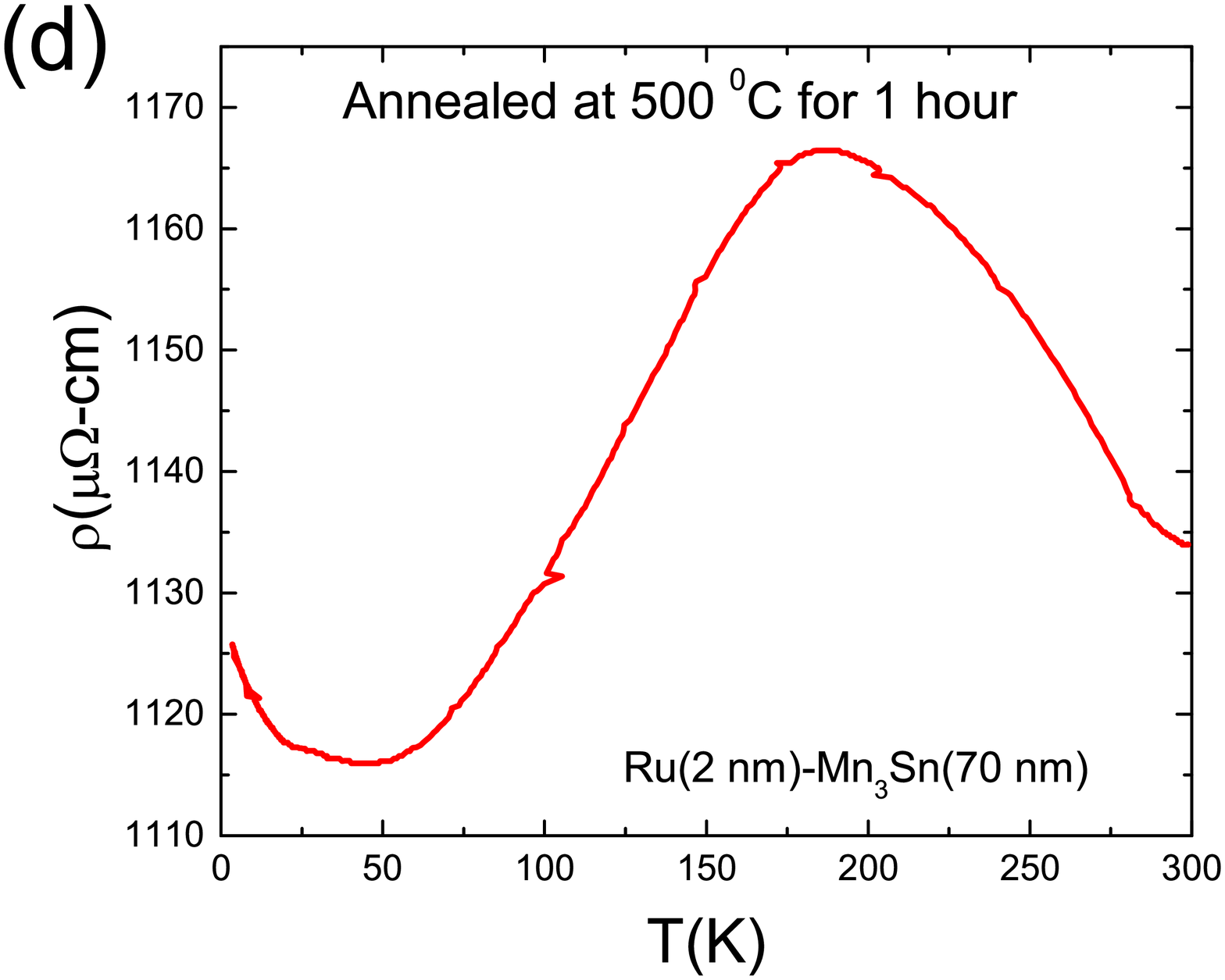}\\

\end{tabular}
\caption{(a) Temperature dependence of resistance of Mn$_3$Sn
nanowire without post annealing. Nano-Hall bar prepared together
with this nanowire did not showed AHE at room temperature. (b,c,d)
Temperature dependence of resistance of different type of
nanowires obtained after annealing at 450 $^0$C and 500 $^0$C for
1 hour. Nano-Hall bars prepared together with this nanowire showed
AHE at room temperature.} \label{FigS2}
\end{figure}
\end{center}

Many of the nanowires were found to be unstable and change from
metallic to semiconducting $R(T)$ during or soon after
measurement. This suggest these nanowires are probably
inhomogeneous and may contain metallic and insulating patches. In
the manuscript we report one typical measurement on the nanowires
which show partly metallic behavior as in Fig. S2(d). The
nano-Hall bar prepared at the same time showed stable metallic
$R(T)$ (as in Fig. S3) and anomalous Hall effect at room
temperature. We would like to emphasize that irrespective of
metallic or semiconducting $R(T)$ all the nanowires showed inverse
spin Hall effect (ISHE) signal at room temperature. We believe the
ISHE signal at room temperature primarily originates from Mn$_3$Sn
phase as Mn-Sn alloy phases have opposite sign of spin Hall angle.
\clearpage

\section{Temperature dependent resistivity of M\lowercase{n$_3$}S\lowercase{n} nano-Hall bar}
\begin{center}
\begin{figure}[h]
\begin{tabular}{ll}
  \centering
  \includegraphics[width= 12 cm]{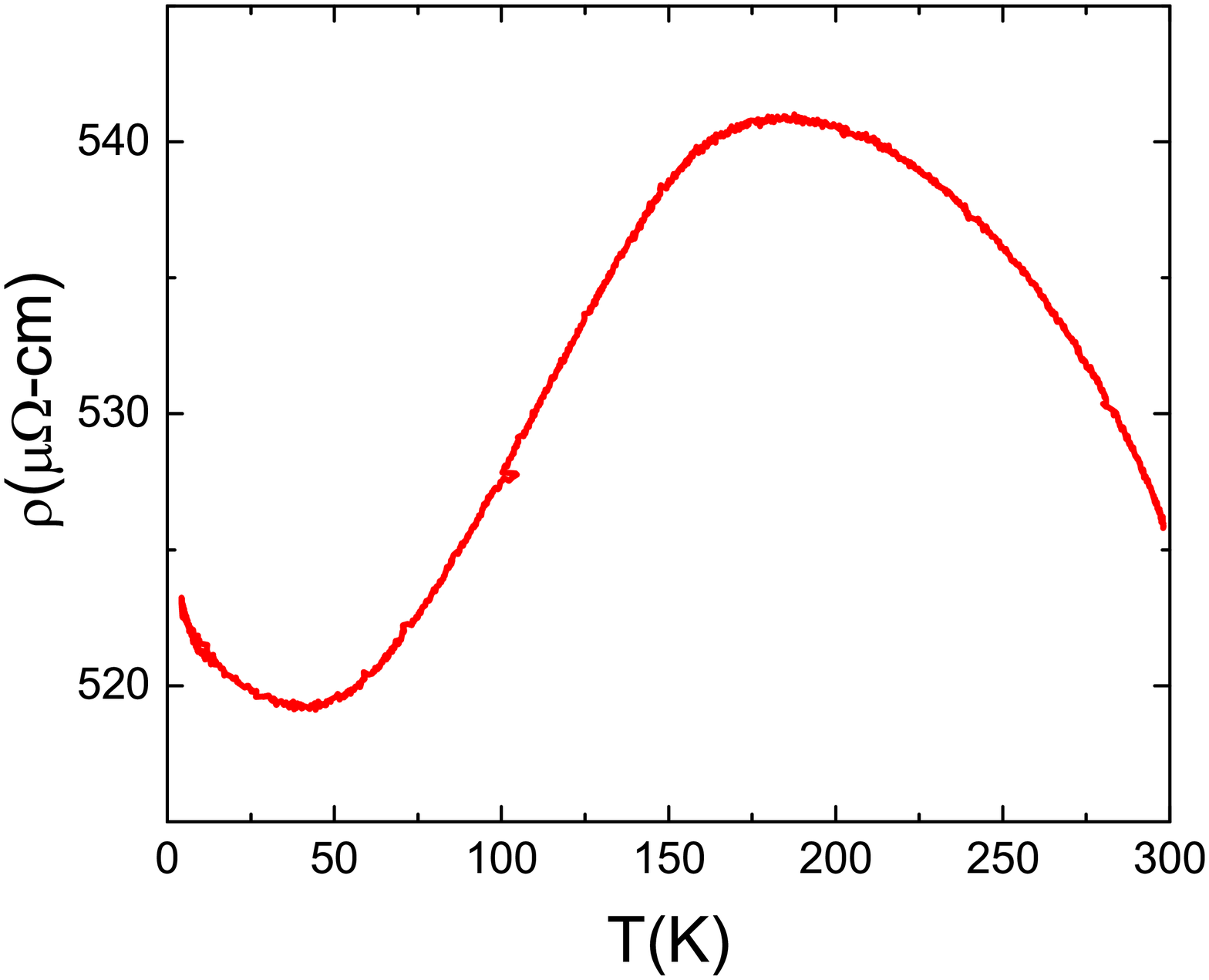}

\end{tabular}
\caption{Shows temperature dependence of resistance of a nano-Hall
bar structure. The resistivity is defined as $\rho  = \frac{V}{I}
\times \frac{{wt}}{L}$, where $V(I)$ is the longitudinal
voltage(current), $w$(= 500 nm) is the width , $L$ (= 3900 nm) is
the length and $t$(= 70 nm) is the thickness of nano-Hall bar.
Note that nano-Hall bar has lower resistivity than nanowires ($w$
= 200 nm) due to larger width ($w$ = 500 nm) of the Hall bar.}
\label{FigS3}
\end{figure}
\end{center}

\newpage

\section{Shunting factor calculation}
\begin{center}
\begin{figure}[h]
\begin{tabular}{ll}
  \centering
  \includegraphics[width= 12 cm]{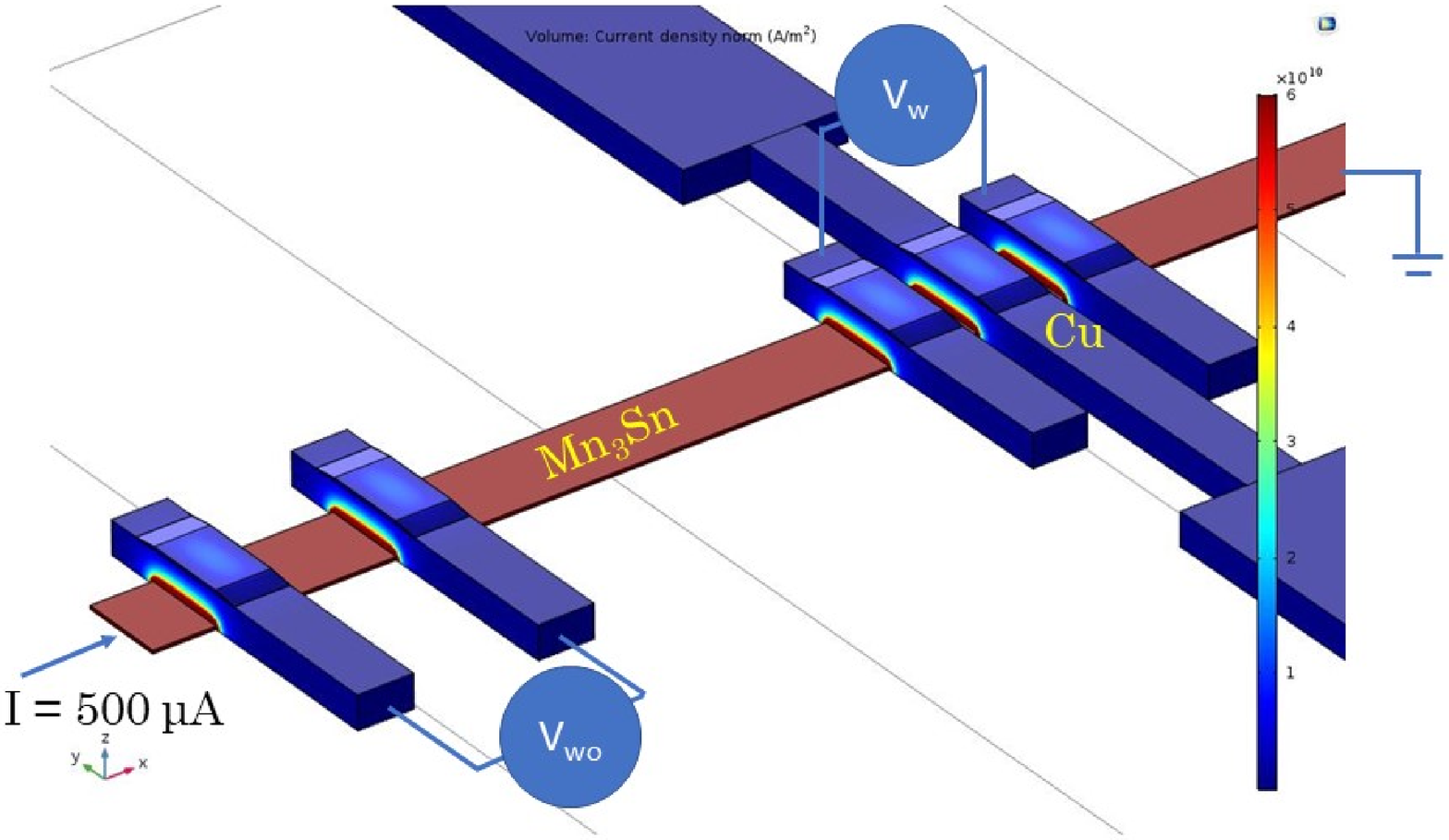}

\end{tabular}
\caption{COMSOL simulation result of current density in the
shunt-device. To determine $x$ , $I$ = 500 $\mu A$ was passed
through the Mn$_3$Sn nanowire and voltages $V_w$  and $V_{wo}$ was
measured from the simulation.} \label{FigS3}
\end{figure}
\end{center}
The spin Hall angle and spin Hall conductivity critically depends
on the shunting factor $x$, which determines amount of charge
current that is shunted back to Cu after spin-charge conversion in
the spin Hall material. In order to estimate $x$ we use a shunt
device (as shown in Fig. S4) that was used previously in [1]. We
use COMSOL AC/DC module to calculate current distribution in this
shunt device. As $x$ critically depends on the width of nanowires
in the device, we measure exact dimension of the spin Hall device
under consideration with SEM and found $w_{Cu}$ =165 nm and
$w_{Mn3Sn}$ =256 nm. The COMSOL simulation was done in a shunting
device with center-to-center distance $L$ =530 nm, $\rho_{Cu}$
=3.72 $\mu\Omega$ cm, and $\rho_{Mn3Sn}$=1133  $\mu\Omega$ cm. The
shunting factor $x$ can be calculated using the equation [1,2],
\begin{equation}
\frac{{V_w }}{{V_{wo} }} = \frac{{L + 2w_{Cu} (x - 1)}}{{L +
w_{Cu} (x - 1)}} = \frac{{40 + 66x}}{{73 + 33x}}
\end{equation}

In this simulation we ignore side shunting by reducing thickness
of spin Hall material to 10 nm. Note that Eq. S1 is valid only for
shunting from top side and is inaccurate in presence of
significant side shunting when thickness of spin Hall material is
comparable to Cu thickness. From COMSOL simulation we found $V_w$
= 0.4748 V and $V_{wo}$ =0.8236 V. Using Eq. S1 we found $x$ =0.04
for our device.  Fig. S5 shows variation of spin Hall angle and
spin Hall conductivity with shunting factor $x$ in the range 0.01
to 0.1.
\begin{center}
\begin{figure}[h]
\begin{tabular}{ll}
  \centering
  \includegraphics[width= 7 cm]{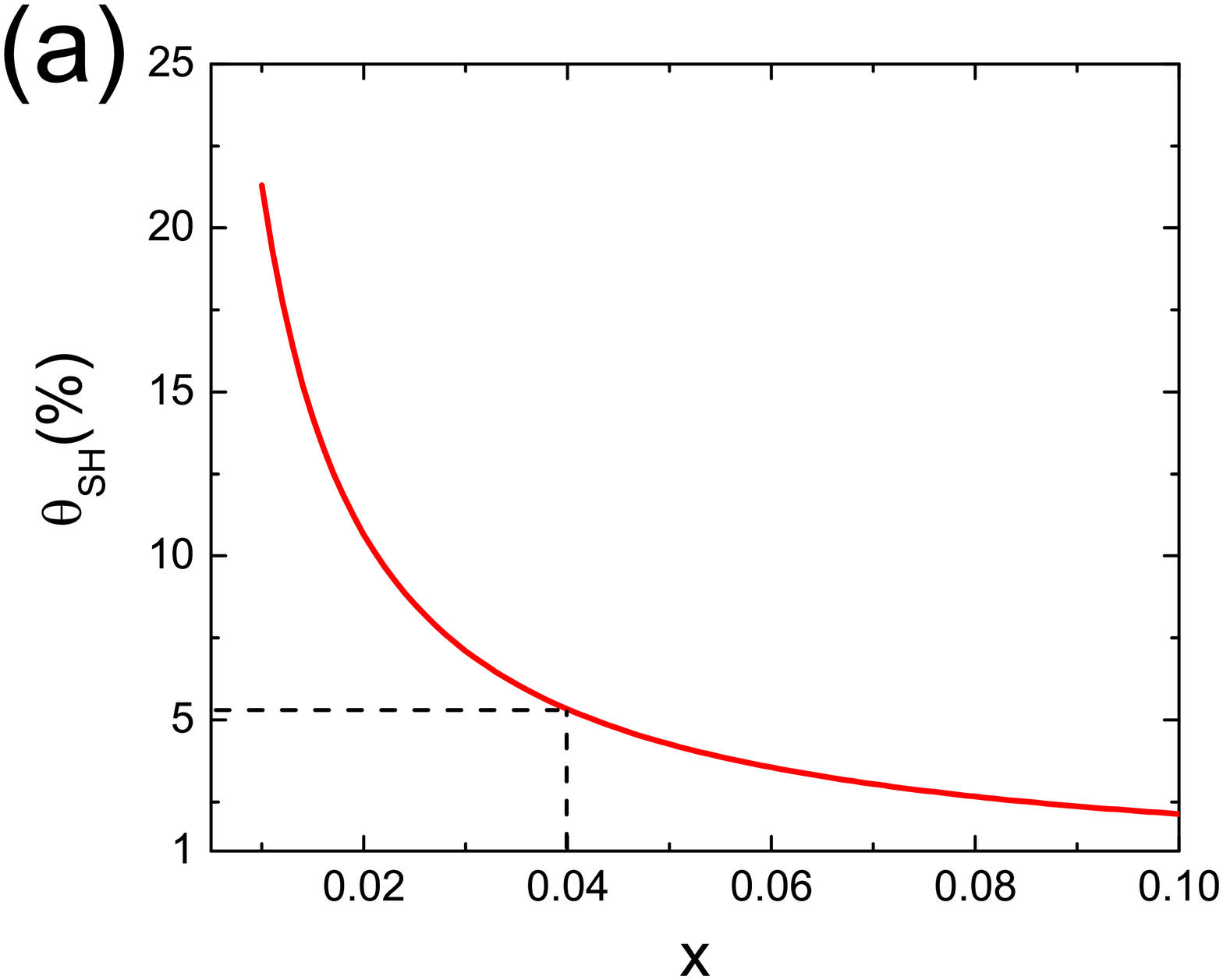}
&

  \includegraphics[width= 7 cm]{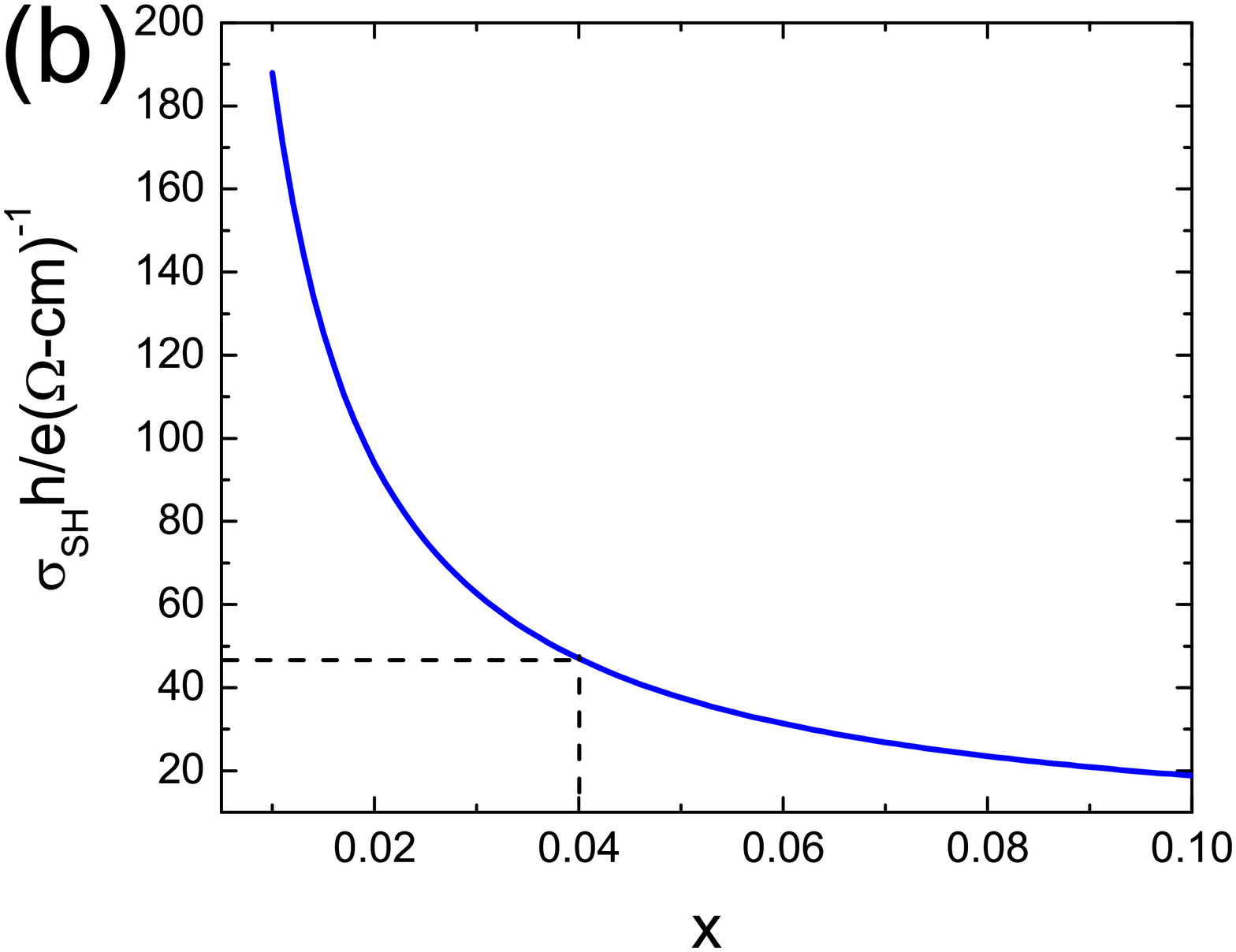}\\

\end{tabular}
\caption{Variation of (a) spin Hall angle $\theta_{SH}$ and (b)
spin Hall conductivity $\sigma_{SH}$  with shunting factor $x$ in
the range 0.01 to 0.1. Note that $x$ = 0.1 is the maximum value of
shunting factor reported [3] experimentally in these devices with
resistivity of spin Hall material of the order of $\sim$900 $\mu
\Omega$- cm.} \label{FigS5}
\end{figure}
\end{center}
\noindent\rule{8cm}{0.4pt}
 .\newline [1] Y. Niimi, M. Morota, D.
H. Wei, C. Deranlot, M. Basletic, A. Hamzic, A. Fert, and Y.
Otani, Phys. Rev. Lett. 106, 126601 (2011).
\newline
[2] PhD Thesis, Miren Isasa Gabilondo, CIC nanoGUNE.
\newline
[3] M. Isasa, M. C. Mart\'{i}nez-Velarte, E. Villamor, C.
Mag\'{o}n, L. Morell\'{o}n, J. M. De Teresa, M. R. Ibarra, G.
Vignale, E. V. Chulkov, E. E. Krasovskii, L. E. Hueso, and F.
Casanova, Phys. Rev. B 93, 014420 (2016).
\newpage

\section{Spin absorption measurement at 10 K.}

\begin{center}
\begin{figure}[h]
\begin{tabular}{ll}
  \centering
  \includegraphics[width= 14 cm]{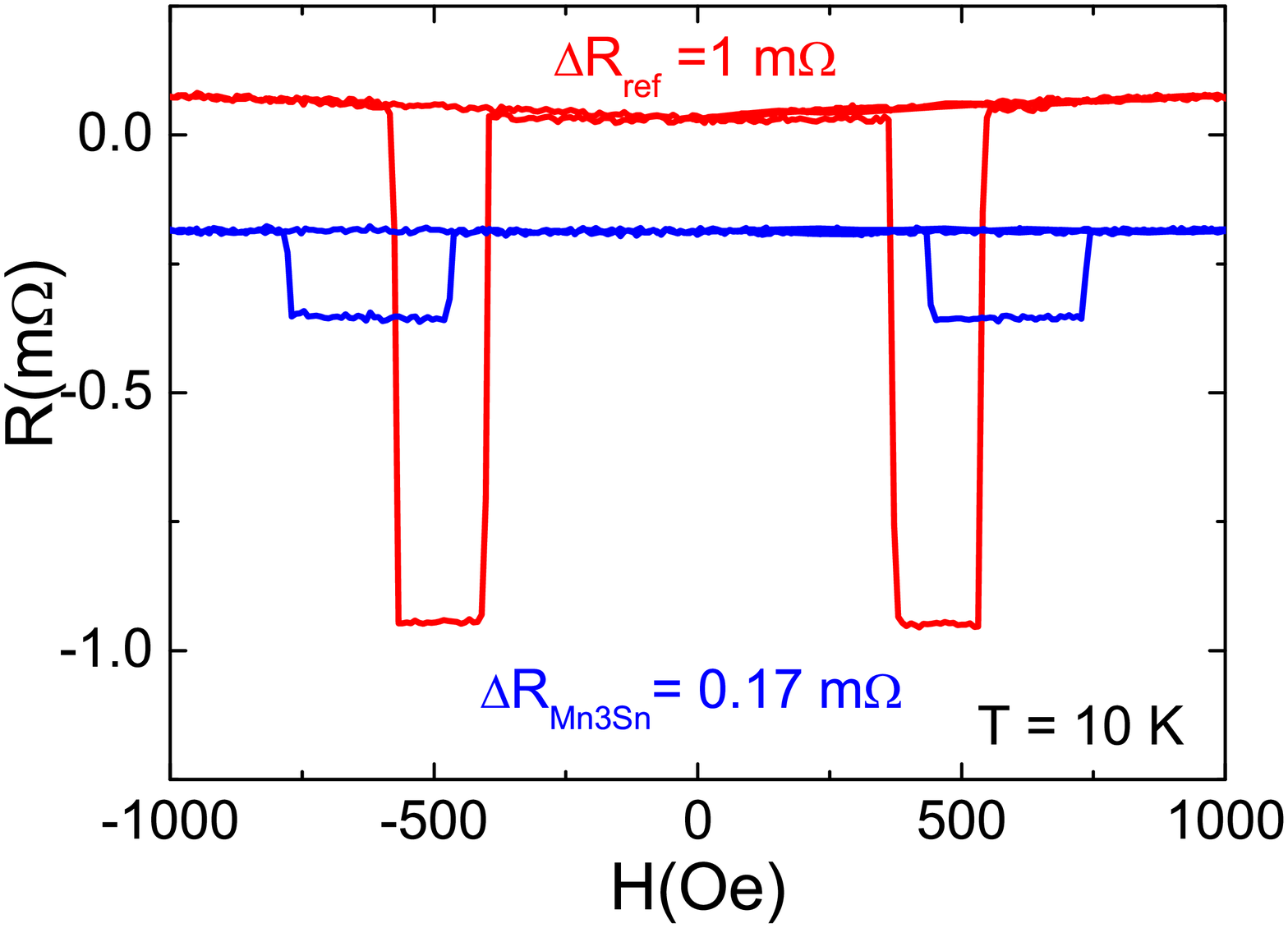}

\end{tabular}
\caption{Shows Nonlocal resistance $R_{NL}$ as a function of
magnetic field for reference (red) and spin Hall device (blue)
measured at 10 K. Measurement done at room temperature on the same
devices is presented in the main text} \label{FigS3}
\end{figure}
\end{center}

\clearpage
\section{Reproducibility}
Most of the freshly prepared nanowires showed inverse spin Hall
effect (ISHE) signal at room temperature. Although magnitude of
ISHE signal was found to vary from device to device.

\begin{center}
\begin{figure}[h]
\begin{tabular}{ll}
  \centering
  \includegraphics[width= 8 cm]{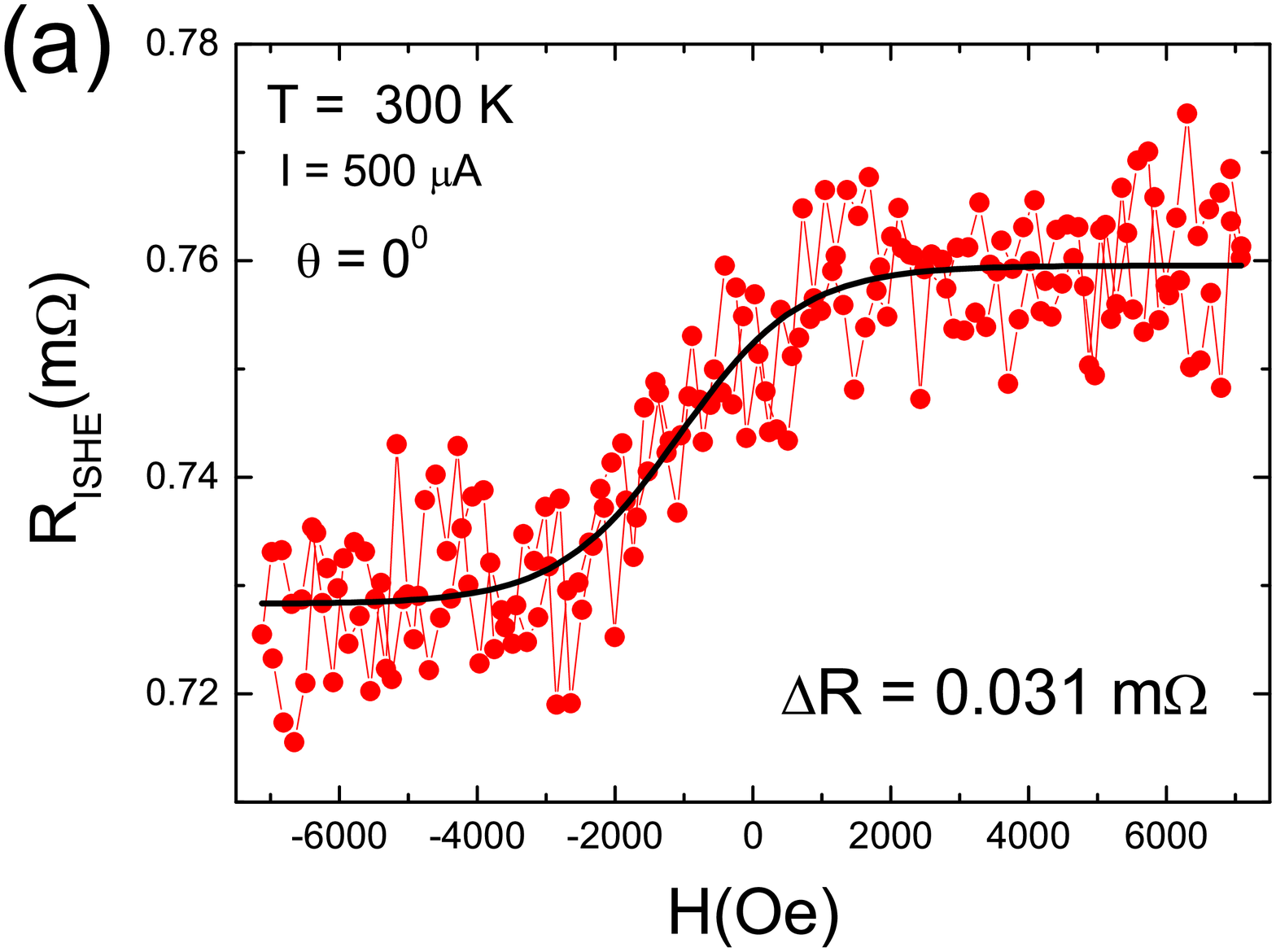}
&

  \includegraphics[width= 8 cm]{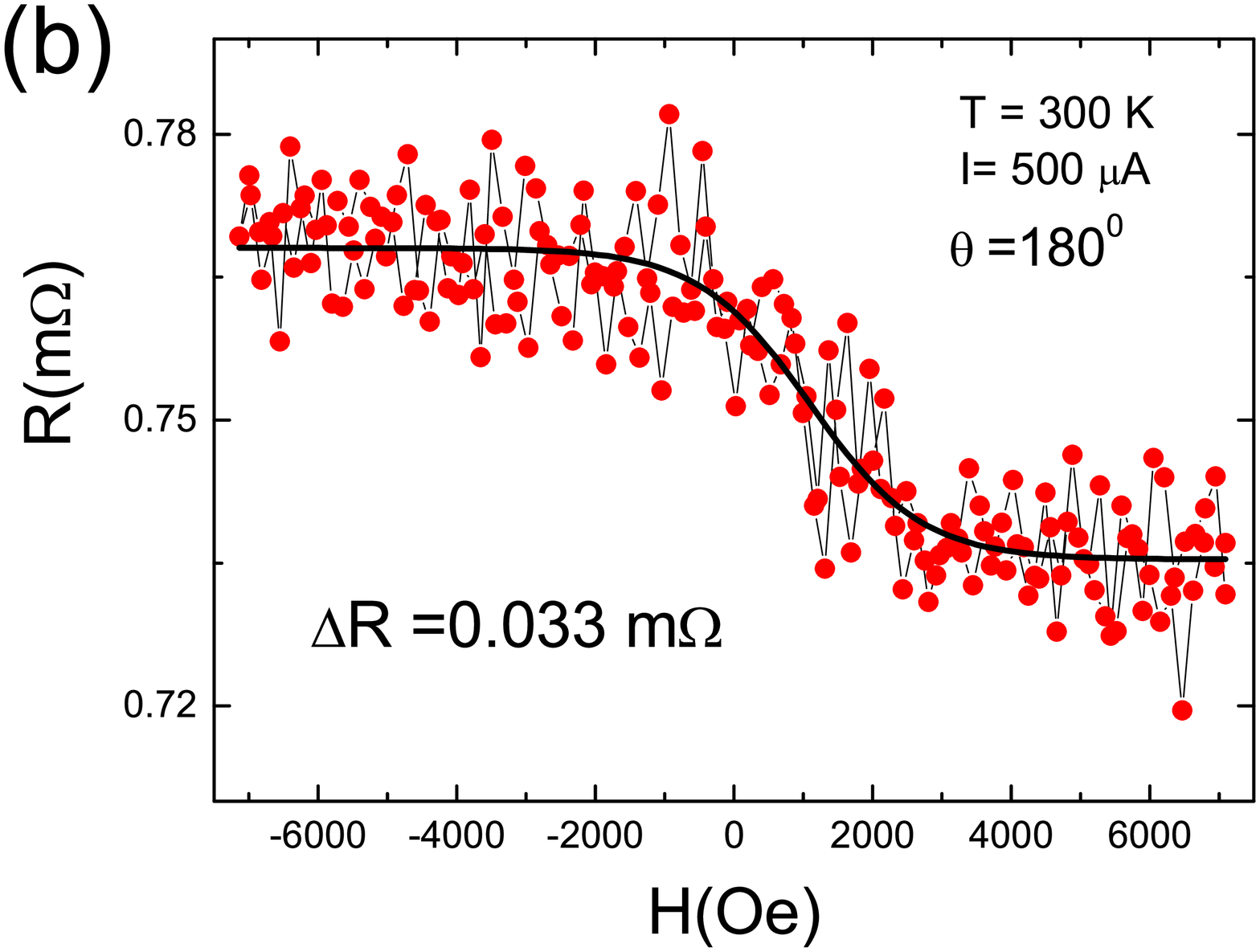}\\

\end{tabular}
\caption{Magnetic field dependence of inverse spin Hall resistance
$R_{ISHE}$ measured at room temperature for magnetic field applied
at (a) $\theta = 0^0$ and (b) $\theta = 180^0$. Magnetic field
angle $\theta = 0^0$ corresponds to field applied in-plane along
Cu spin transport channel which is  perpendicular to Mn$_3$Sn
nanowire (direction as shown in Fig. 2(a)). This ISHE signal
measurement is obtained in the semiconducting Mn$_3$Sn nanowire
shown in Fig. S2(c). } \label{FigS7}
\end{figure}
\end{center}

\begin{center}
\begin{figure}[h]
\begin{tabular}{ll}
  \centering
  \includegraphics[width= 12 cm]{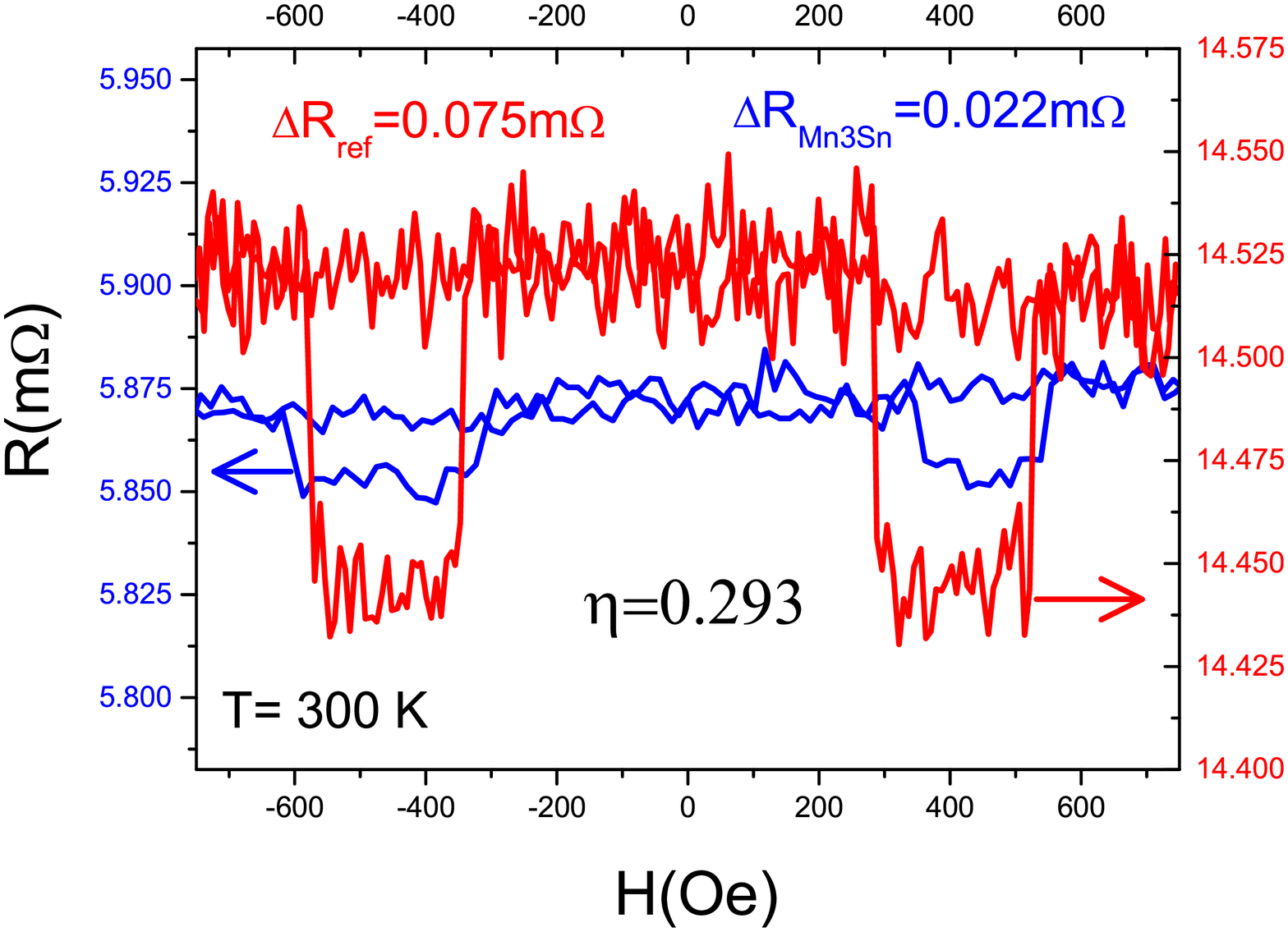}

\end{tabular}
\caption{Nonlocal resistance $R_{NL}$ as a function of magnetic
field for reference and spin Hall device measured at room
temperature. This measurement was done on the same semiconducting
Mn$_3$Sn nanowire as shown in Fig. S7. We found spin diffusion
length $\lambda_S$ $\sim$0.46 nm in this pair of devices
comparable to $\lambda_S$ of the device reported in the main
text.} \label{FigS8}
\end{figure}
\end{center}

\end{widetext}


\begin{thebibliography}{100}

\bibitem{baltz}
V. Baltz, A. Manchon, M. Tsoi, T. Moriyama, T. Ono, and Y.
Tserkovnyak, Rev. Mod. Phys. \textbf{90}, 015005 (2018).

\bibitem{smejjal}
L. \v{S}mejkal, T. Jungwirth, and J. Sinova, Phys. Status Solidi
RRL \textbf{11}, 1700044 (2017).

\bibitem{smejjal-natphy}
L. \v{S}mejkal,  Y. Mokrousov, B. Yan, and A. H. MacDonald, Nat.
Phys. \textbf{14}, 242 (2018).

\bibitem{Barthem}
V. M. T. S. Barthem, C. V. Colin, H. Mayaffre, M.-H. Julien, and
D. Givord, Nat. Commun. \textbf{4}, 2892 (2013).

\bibitem{wadley}
P. Wadley, B. Howells, J. Zelezny, C. Andrews, V. Hills, R. P.
Campion, V. Novak, K. Olejn\'{i}k, F. Maccherozzi, S. S. Dhesi, S.
Y. Martin, T. Wagner, J. Wunderlich, F. Freimuth, Y. Mokrousov, J.
Kunes, J. S. Chauhan, M. J. Grzybowski, A. W. Rushforth, K. W.
Edmonds, B. L. Gallagher, and T. Jungwirth, Science \textbf{351},
587 (2016).

\bibitem{Zelezny-natphy}
J. \u{Z}elezn\'{y}, P. Wadley, K. Olejnik, A. Hoffmann, and H.
Ohno, Nat. Phys. \textbf{14}, 220 (2018).

\bibitem{Park}
J. Park, G. Lee, F. Wolff-Fabris, Y. Y. Koh, M. J. Eom, Y. K. Kim,
M. A. Farhan, Y. J. Jo, C. Kim, J. H. Shim, and J. S. Kim, Phys.
Rev. Lett. \textbf{107}, 126402 (2011).

\bibitem{Masuda}
H. Masuda, H. Sakai, M. Tokunaga, Y. Yamasaki, A. Miyake, J.
Shiogai, S. Nakamura, S. Awaji, A. Tsukazaki, H. Nakao, Y.
Murakami, T.-h. Arima, Y. Tokura, and S. Ishiwata, Sci. Adv.
\textbf{2}, e1501117 (2016).

\bibitem{Richard}
P. Richard, K. Nakayama, T. Sato, M. Neupane, Y.-M. Xu, J. H.
Bowen, G. F. Chen, J. L. Luo, N. L. Wang, X. Dai, Z. Fang, H.
Ding, and T. Takahashi, Phys. Rev. Lett. \textbf{104}, 137001
(2010).

\bibitem{Awang}
A. Wang, I. Zaliznyak, W. Ren, L. Wu, D. Graf, V. O. Garlea, J. B.
Warren, E. Bozin, Y. Zhu, and C. Petrovic, Phys. Rev. B
\textbf{94}, 165161 (2016).

\bibitem{muller}
R. A. M\"{u}ller, N. R. Lee-Hone, L. Lapointe, D. H. Ryan, T.
Pereg-Barnea, A. D. Bianchi, Y. Mozharivskyj, and R. Flacau, Phys.
Rev. B \textbf{90}, 041109(R) (2014).


\bibitem{Hirschberger}
M. Hirschberger, S. Kushwaha, Z. Wang, Q. Gibson, S. Liang, C. A.
Belvin, B. A. Bernevig, R. J. Cava, and N. P. Ong, Nat. Mater.
\textbf{15}, 1161 (2016).

\bibitem{ZFwang}
Z. F. Wang, H. Zhang, D. Liu, C. Liu, C. Tang, C. Song, Y. Zhong,
J. Peng, F. Li, C. Nie, L. Wang, X. J. Zhou, X. Ma, Q. K. Xue, and
F. Liu, Nat. Mater. \textbf{15}, 968 (2016).

\bibitem{Wakeham}
N. Wakeham, E. D. Bauer, M. Neupane, and F. Ronning, Phys. Rev. B
\textbf{93}, 205152 (2016).

\bibitem{Sushkov}
A. B. Sushkov, J. B. Hofmann, G. S. Jenkins, J. Ishikawa, S.
Nakatsuji, S. Das Sarma, and H. D. Drew, Phys. Rev. B \textbf{92},
241108(R) (2015).

\bibitem{chen}
H. Chen, Q. Niu, and A. H. MacDonald, Phys. Rev. Lett.
\textbf{112}, 017205 (2014).

\bibitem{nakatsuji}
S. Nakatsuji, N. Kiyohara, and T. Higo, Nature (London)
\textbf{527}, 212 (2015).

\bibitem{nayak}
A. K. Nayak, J. E. Fischer, Y. Sun, B. Yan, J. Karel, A. C.
Komarek, C. Shekhar, N. Kumar, W. Schnelle, J. K\"{u}bler, C.
Felser, and S. S. P. Parkin, Sci. Adv. \textbf{2}, e1501870
(2016).


\bibitem{manna}
K. Manna, Y. Sun, L. Müchler, J. K\"{u}bler, and C. Felser, Nat.
Rev. Mater. \textbf{3}, 244 (2018).

\bibitem{Kiyohara}
N. Kiyohara, T. Tomita, and S. Nakatsuji, Phys. Rev. Appl.
\textbf{5}, 064009 (2016).

\bibitem{kubler-el}
J. K\"{u}bler, C. Felser, Europhys. Lett. \textbf{108}, 67001
(2014).

\bibitem{kubler}
J. K\"{u}bler and C. Felser, Europhys. Lett. \textbf{120}, 47002
(2017).

\bibitem{yang}
H. Yang, Y. Sun, Y. Zhang, W.-J. Shi, S. S. P. Parkin, and B. Yan,
New J. Phys. 19, 015008 (2017).

\bibitem{Tian}
S. Nakatsuji, T. Higo, M. Ikhlas, T. Tomita, and Z. Tian, Philos.
Mag. \textbf{97}, 2815 (2017).

\bibitem{Brown}
P. J. Brown, V. Nunez, F. Tasset, J. B. Forsyth, and P.
Radhakrishna, J. Phys.: Condens. Matter \textbf{2}, 9409 (1990).


\bibitem{Nagamiya}
T. Nagamiya, S. Tomiyoshi, Y. Yamaguchi, Solid State
Commun.\textbf{ 42}, 385.388 (1982).

\bibitem{Guo-prb}
G. Guo and T. Wang, Phys. Rev. B \textbf{96}, 224415 (2017).


\bibitem{kuroda}
K. Kuroda, T. Tomita, M.-T. Suzuki, C. Bareille, A. A. Nugroho, P.
Goswami, M. Ochi, M. Ikhlas, M. Nakayama, S. Akebi, R. Noguchi, R.
Ishii, N. Inami, K. Ono, H. Kumigashira, A. Varykhalov, T. Muro,
T. Koretsune, R. Arita, S. Shin, T. Kondo, and S. Nakatsuji, Nat.
Mater \textbf{16}, 1090 (2017).

\bibitem{Xiaokang}
X. Li, L. Xu, L. Ding, J. Wang, M. Shen, X. Lu, Z. Zhu, and K.
Behnia, Phys. Rev. Lett. \textbf{119}, 056601 (2017).


\bibitem{narita}
H. Narita, M. Ikhlas, M. Kimata, A. A. Nugroho, S. Nakatsuji, and
Y. Otani, Appl. Phys. Lett. \textbf{111}, 202404 (2017).


\bibitem{Ikhlas}
M. Ikhlas, T. Tomita, T. Koretsune, M.-T. Suzuki, D.
Nishio-Hamane, R. Arita, Y. Otani, and S. Nakatsuji, Nat. Phys 13,
1085 (2017).

\bibitem{Xiaokang-scipost}
X. Li, L. Xu, H. Zuo, A. Subedi, Z. Zhu, K. Behnia, SciPost Phys.
5, 063 (2018).


\bibitem{Higo-nat-photon}
T. Higo, H. Man, D. B. Gopman, L. Wu, T. Koretsune, O. M. J. van't
Erve, Y. P. Kabanov, D. Rees, Y. Li, M.-T. Suzuki, S. Patankar, M.
Ikhlas, C. L. Chien, R. Arita, R. D. Shull, J. Orenstein, and S.
Nakatsuji, Nat. Photonics \textbf{12}, 73 (2018).


\bibitem{Markou}
A. Markou, J. M. Taylor, A. Kalache, P. Werner, S. S. P. Parkin,
and C. Felser, Phys. Rev. Mater. \textbf{2}, 051001(R) (2018).


\bibitem{THigo}
T. Higo, D. Qu, Y.Li, C. L. Chien, Y. Otani and S. Nakatsuji,,
Appl. Phys. Lett. \textbf{113}, 202402 (2018).

\bibitem{you}
Y. You, X. Chen, X. Zhou, Y. Gu, R. Zhang, F. Pan, C. Song, Adv.
Electron. Mater. 1800818 (2019).


\bibitem{Guo}
G. Y. Guo, S. Murakami, T.-W. Chen, and N. Nagaosa, Phys. Rev.
Lett. \textbf{100}, 096401 (2008).

\bibitem{Nagaosa-rev}
N. Nagaosa, J. Sinova, S. Onoda, A. H. MacDonald, and N. P. Ong,
Rev. Mod. Phys. 82, 1539 (2010).

\bibitem{Jungwirth}
T. Jungwirth, Q. Niu, and A. H. MacDonald, Phys. Rev. Lett.
\textbf{88}, 207208, (2002).


\bibitem{omori}
Y. Omori, E. Sagasta, Y. Niimi, M. Gradhand, L-E. Hueso, F.
Casanova and Y. Otani,Phys. Rev. B \textbf{99}, 014403 (2019).

\bibitem{Sun}
Y. Sun, Y. Zhang, C. Felser, and B. Yan, Phys. Rev. Lett.
\textbf{117}, 146403 (2016).


\bibitem{Zhang-prb}
Y. Zhang, Y. Sun, H. Yang, J. \u{Z}elezn\'{y}, S. P. P. Parkin, C.
Felser, and B. Yan, Phys. Rev. B \textbf{95}, 075128 (2017).



\bibitem{zhang}
Y. Zhang, J. \u{Z}elezn\'{y}, Y. Sun, J. van den Brink, and B.
Yan, New J. Phys. \textbf{20},073028 (2018).


\bibitem{Zhang-sciadv}
W. Zhang, W. Han, S.-H. Yang, Y. Sun, Y. Zhang, B. Yan, and S. S.
P. Parkin, Sci. Adv. \textbf{2}, e1600759 (2016).


\bibitem{Jakub}
J. \u{Z}elezn\'{y}, Y. Zhang, C. Felser, and B. Yan, Phys. Rev.
Lett. \textbf{119}, 187204 (2017).


\bibitem{huachen}
H. Chen, Q. Niu, and A. H. MacDonald, arXiv:1803.01294v1


\bibitem{supp}
See Supplemental Material at http://link.aps.org/supplemental/ for
details of the nanowire fabrication process, x-ray diffraction of
Mn$_3$Sn thin film, RT of nano-Hall bar,  calculation of shunting
factor $x_{sh}$ by COMSOL software, spin absorption measurement at
10 K and reproducibility data.

\bibitem{Filippou}
P. C. Filippou, J. Jeong, Y. Ferrante, S.-H. Yang, T. Topuria, M.
G. Samant and S. S. P. Parkin, Nat. Commun. \textbf{9}, 4653
(2018).


\bibitem{wen}
Z. C. Wen, J. Kim, H. Sukegawa, M. Hayashi, and S. Mitani, AIP
Adv. \textbf{6}, 056307 (2016).


\bibitem{Zimmer}
G. J. Zimmer and E. Kr\'{e}n, AIP Conf. Proc. \textbf{5}, 513
(1972); E. Kren, J. Paitz, G. Zimmer, and \'{E}. Zsoldos, Physica
B \textbf{80}, 226 (1975).


\bibitem{Sung}
N. H. Sung, F. Ronning, J. D. Thompson, and E. D. Bauer, Appl.
Phys. Lett. \textbf{112},132406 (2018).


\bibitem{Fuchs}
K. Fuchs, Proc. Cambridge Philos. Soc. \textbf{34}, 100 (1938); E.
H. Sondheimer, Adv. Phys. \textbf{1}, 1 (1952); A. F. Mayadas and
M. Shatzkes, Phys. Rev. B \textbf{1}, 1382 (1970).


\bibitem{edurne}
E. Sagasta, Y. Omori, M. Isasa, M. Gradhand, L. E. Hueso, Y.
Niimi, Y. Otani, and F. Casanova, Phys. Rev. B \textbf{94},
060412(R) (2016).


\bibitem{Sagasta}
E. Sagasta, Y. Omori, S. V\'{e}ez, R. Llopis, C. Tollan, A.
Chuvilin, L. E. Hueso, M. Gradhand, Y. Otani, and F. Casanova,
Phys. Rev. B \textbf{98}, 060410(R) (2018).



\bibitem{muduli}
P. K. Muduli, M. Kimata, Y. Omori, T. Wakamura, S. P. Dash, Y. C.
Otani, Phys. Rev. B \textbf{98}, 024416 (2018).

\bibitem{WZhang-prl}
W. Zhang, M. B. Jungfleisch, W. Jiang, J. E. Pearson, A. Hoffmann,
F. Freimuth, and Y. Mokrousov, Phys. Rev. Lett. \textbf{113},
196602 (2014).

\bibitem{Morota}
M. Morota, Y. Niimi, K. Ohnishi, D. H. Wei, T. Tanaka, H. Kontani,
T. Kimura and Y. Otani, Phys. Rev. B \textbf{83}, 174405 (2011).

\bibitem{niimi}
Y. Niimi, and Y. Otani, Rep. Prog. Phys. \textbf{78}, 124501
(2015).

\bibitem{note1}
The resistivity value of Mn$_3$Sn nanowire used in the calculation
is $\rho_{Mn_3Sn}$ $\approx$1133 $\mu\Omega$ cm. If the calculaton
is replicated for bulk resistivity $\rho_{Mn_3Sn}$ $\approx$320
$\mu\Omega$ cm,  we obtain  $\theta_{SH}$ $\sim$5.34 $\%$ , spin
diffusion length $\lambda_{s(Mn_3Sn)}$ $\sim$2.64  nm and spin
Hall conductivity $\sigma_{SH}$ $\sim$167.06 ($\hbar/e$) ($\Omega$
cm)$^{-1}$. Note that spin Hall angle remains unchanged due to
increase in $\lambda_{s(Mn_3Sn)}$  for lower $\rho_{Mn_3Sn}$.

\bibitem{Qu}
D. Qu, T. Higo, T. Nishikawa, K. Matsumoto, K. Kondou, D.
Nishio-Hamane, R. Ishii, P. K. Muduli, Y. Otani, and S. Nakatsuji,
Phys. Rev. Materials \textbf{2}, 102001(R) (2018).

\bibitem{Kimata}
M. Kimata, H. Chen, K. Kondou, S. Sugimoto, P. K. Muduli, M.
Ikhlas, Y. Omori, T. Tomita, A. H. MacDonald, Satoru Nakatsuji,
YoshiChika Otani, Nature \textbf{565}, 627 (2019).

\bibitem{Liu}
Z. Q. Liu, H. Chen, J. M. Wang, J. H. Liu, K. Wang, Z. X. Feng, H.
Yan, X. R. Wang, C. B. Jiang, J. M. D. Coey and A. H. MacDonald,
Nat. Electron. \textbf{1}, 172 (2018).


\end{thebibliography}
\end{document}